\documentclass[amsthm]{elsart}
\usepackage{amsmath,amssymb,url}
\usepackage{graphicx,msc}
\usepackage{cite}
\usepackage{color}
\usepackage{xspace}
\usepackage{pstricks}

\renewcommand{\note}[2]{

\noindent {\blue {\sf {\bf [#1]} #2} }

}
\renewcommand{\note}[2]{}

\newcommand{\z}[1]{\mathit{#1}}
\newcommand{\zt}[1]{\textit{#1}}
\newcommand{\powerset}[1]{{\cal P(#1)}}         %Powerset of a set
\newcommand{\domain}{\z{dom}}                   %Domain function
\newcommand{\image}{\z{im}}
\newcommand{\anything}{\cdot}                   %Just a placeholder
\newcommand{\equivalent}{\iff}                  %Symbol for equivalence
\newcommand{\defeq}{\stackrel{\mathrm{def}}{=}} %Definition equation

\newcommand{\roletermset}{\z{RoleTerm}}        %Set of role terms
\newcommand{\funcset}{{\cal F}}                %Set of function names
\newcommand{\labelset}{{\cal L}}               %Set of labels
\newcommand{\thelabel}{\ell}                   %Just a label called l
\newcommand{\idset}{{\cal ID}}                 %Set of identifiers
\newcommand{\roleset}{{\cal R}}                %Set of roles
\newcommand{\eventset}{{\cal E}}               %Set of events
\newcommand{\idtyper}{\z{type}}            %Function: role -> {id -> { const, param, var }}
\newcommand{\consttype}{{\z{const}}}           %Type of constant id
\newcommand{\paramtype}{{\z{param}}}           %Type of parameter id
\newcommand{\vartype}{{\z{variable}}}          %Type of variable id
\newcommand{\varterm}[2]{\z{var}_{#1}(#2)}     %Set of variables in term for a given role

\newcommand{\sessionuniqueclaim}{\z{session\text{-}unique}}
\newcommand{\dataagreeclaim}{\z{data\text{-}agree}}
\newcommand{\sessionclaim}{\z{session}}        %Claim of session property
\newcommand{\wsessionclaim}{\z{wsession}}      %Claim of weak session property
\newcommand{\secretclaim}{\z{secret}}          %Claim of secrecy
\newcommand{\nisynchclaim}{\z{synch}}          %Claim of non-inj synchr.
\newcommand{\isynchclaim}{\z{i\text{-}synch}}       %Claim of inj synchr.

\newcommand{\claimset}{\z{Claim}}              %Set of claims
\newcommand{\rs}{\mathit{rs}}                  %Concrete role specification
\newcommand{\elist}{\mathit{elist}}
\newcommand{\rolespec}{\z{RoleSpec}}           %Role specification set
\newcommand{\createevent}{\z{create}}          %Create event
\newcommand{\readevent}{\z{read}}              %Read event
\newcommand{\sendevent}{\z{send}}              %Send event
\newcommand{\claimevent}{\z{claim}}            %Claim event
\newcommand{\stopevent}{\z{end}}               %End event
\newcommand{\noevent}{\varepsilon}             %No event
\newcommand{\before}{\prec}                    %Ordering on events
\newcommand{\seq}{\cdot}                       %Constructing lists of events

\newcommand{\protocols}{\z{Prot}}              %Universe of protocols

\newcommand{\runtermset}{\z{RunTerm}}          %Set of run terms
\newcommand{\runidset}{\z{Runid}}              %Set of run identifiers
\newcommand{\Inst}{\z{Inst}}                   %Set of instantiations
\newcommand{\run}{\z{Runs}}                    %Set of runs
\newcommand{\agentset}{{\cal A}}               %Set of run agents
\newcommand{\trustedagentset}{\agentset_T}     %Set of trusted agents
\newcommand{\untrustedagentset}{\agentset_U}     %Set of trusted agents
\newcommand{\undef}{\bot}                      %Undefined value for inst

\newcommand{\markid}[1]{\sharp{#1}}            %Run ident mark

\newcommand{\IK}{M}                            %Intruder knowledge set
\newcommand{\IT}{\mathcal{IT}}                 %Intruder-generated terms.

\newcommand{\traces}{\z{Traces}}               %Set of traces
\newcommand{\tracesof}[1]{\z{Tr}(#1)}
\newcommand{\restrictedtraces}[2]{\tracesof{#1;#2}}

\newcommand{\transition}[3]{#1 \shortarrow{#2} #3}
\newcommand{\shortarrow}[1]{\stackrel{#1}{\rightarrow}}
\newcommand{\sosrule}[3]{{[\z{#1}]}\frac{#2}{#3}}

\newcommand{\subterm}{\sqsubseteq}
\newcommand{\inv}{\ensuremath{^{-1}}}          %Inverse key function
\newcommand{\lenc}{\{\mkern-4.0mu\lvert\,}     % Left encryption brace
\newcommand{\renc}{\,\rvert\mkern-4.0mu\}}     % Right encryption brace
\newcommand{\enc}[2]{ \lenc {#1} \renc_{#2}}   %encryption
\newcommand{\closure}[1]{\overline{#1}}        %Closure of intruder knowledge
\newcommand{\runsof}{\z{runsof}}               %Runs of function
\newcommand{\runids}{\z{runids}}               %Run identifiers function
\newcommand{\length}[1]{|#1|}                  %Length of a trace
\newcommand{\Match}{\z{Match}}                 %Unification predicate
\newcommand{\cast}{\z{cast}}                    %cast: mapping of roles to runids
\newcommand{\castset}[2]{\z{Cast}(#1,#2)} % Casts of protocol #1 in trace #2.

\newcommand{\satisfies}[2]{\mathrm{sat}(#1,#2)}
\newcommand{\restrictedsatisfies}[3]{\mathrm{sat}(#1,#2;#3)}

\newcommand{\adversaryinput}{\z{adversary}}
\newcommand{\ioout}[2]{#1\langle #2 \rangle} % #1:protocol (set), #2:output
\newcommand{\iooutreveal}[2]{#1 \langle #2^* \rangle}
\newcommand{\ioin}[3]{#1\langle #2/#3 \rangle} % #1:protocol (set),
\newcommand{\ioinunique}[3]{#1\langle #2!/#3 \rangle} % #1:protocol (set),
\newcommand{\ioinadversary}[2]{\ioin{#1}{\adversaryinput}{#2}}
\newcommand{\ioinsecret}[2]{\ioin{#1}{\secretclaim}{#2}}
\newcommand{\ioinsession}[2]{\ioin{#1}{\sessionclaim}{#2}}

\newcommand{\independent}[2]{\mathrm{indep}(#1,#2)}
\newcommand{\independentc}[3]{\mathrm{indep}(#1,#2;#3)} % In context #3
\newcommand{\sindependent}[2]{\mathrm{s\text{-}indep}(#1,#2)}
\newcommand{\nonce}[1]{n#1}                 %Nonce
\newcommand{\publickey}[1]{pk(#1)}          %Public key
\newcommand{\secretkey}[1]{sk(#1)}          %Secret key
\newcommand{\noni}{\nonce{i}}               %Nonce i
\newcommand{\nonr}{\nonce{r}}               %Nonce r
\newcommand{\nonip}{\nonce{i'}}             %Nonce i'
\newcommand{\nonrp}{\nonce{r'}}             %Nonce r'
\newcommand{\pki}{\publickey{i}}            %Public key i
\newcommand{\pkr}{\publickey{r}}            %Public key r
\newcommand{\ski}{\secretkey{i}}            %Secret key i
\newcommand{\irole}{i}                      %Name of i
\newcommand{\rrole}{r}                      %Name of r

\newcommand{\chaincomp}[4]{#1\cdot #2}   % Chaining composition

\newcommand{\wimax}{WiMAX\xspace}
\newcommand{\nsl}{\ensuremath{\mathit{NSL}}}

\newcommand{\ranging}{\zt{Ranging}}           %The ranging protocol
\newcommand{\registration}{\zt{Registration}} %The registration protocol
\newcommand{\pkmvrsa}{\zt{PKMv2 RSA}}         %The PKMv2-RSA protocol
\newcommand{\pkmvsatek}{\zt{PKMv2 SA-TEK}}    %The PKMv2 SA-TEK protocol
\newcommand{\pkmvtek}{\zt{PKMv2 Key}}  %The PKMv2 TEK update
\newcommand{\ms}{\zt{ms}}            %Name of the Mobile Station
\newcommand{\bs}{\zt{bs}}            %Name of the Base Station
\newcommand{\man}{\zt{man}}          %Name of the Mobile Station's manufacturer
\newcommand{\ca}{\zt{ca}}          %Name of some certification authority
\newcommand{\bcid}{\zt{BCID}}        %Basic Connection Identifier
\newcommand{\skms}{\secretkey{\ms}}         %Secret key of mobile station
\newcommand{\skman}{\secretkey{\man}}       %Secret key of manufacturer
\newcommand{\skbs}{\secretkey{\bs}}         %Secret key of base station
\newcommand{\skca}{\secretkey{\ca}}         %Secret key of cert auth
\newcommand{\pkms}{\publickey{\ms}}         %Public key of mobile station
\newcommand{\pkbs}{\publickey{\bs}}         %Public key of base station
\newcommand{\pkca}{\publickey{\ca}}         %Public key of cert auth

\newcommand{\authinfo}{\zt{Auth Info}}   %AUTH-INFO message
\newcommand{\rsareq}{\zt{Request}}       %RSA-REQ message
\newcommand{\rsareply}{\zt{Reply}}   %RSA-REPLY message
\newcommand{\rsareject}{\zt{Reject}} %RSA-REJECT message
\newcommand{\rsaack}{\zt{Acknowledgment}}       %RSA-ACK message

\newcommand{\macert}{\zt{MAN-Crt}}       %manufacturer's certificate
\newcommand{\mscert}{\zt{MS-Crt}}       %mobile station's certificate
\newcommand{\bscert}{\zt{BS-Crt}}       %base station's certificate
\newcommand{\msrand}{\zt{MS-Rnd}}       %mobile station's nonce
\newcommand{\bsrand}{\zt{BS-Rnd}}       %base station's nonce
\newcommand{\said}{\zt{SAID}}            %Security Association Id.
\newcommand{\prepak}{\zt{pPAK}}       %pre-PAK

\newcommand{\satekchall}{\zt{Challenge}}   %SA-TEK-challenge message
\newcommand{\satekrequest}{\zt{Request}}   %SA-TEK-request message
\newcommand{\satekresponse}{\zt{Response}}   %SA-TEK-response message

\newcommand{\ak}{\zt{AK}}            %Authentication key
\newcommand{\pak}{\zt{PAK}}          %Primary authentication key
\newcommand{\kek}{\zt{KEK}}              %Key encryption key
\newcommand{\tek}{\zt{TEK}}              %Traffic encryption key
\newcommand{\tekzero}{\ensuremath{\tek_0}}           %Old traffic encryption key
\newcommand{\tekone}{\ensuremath{\tek_1}}            %New traffic encryption key
\newcommand{\aksn}{\zt{AKSN}}        %AK's sequence number
\newcommand{\akid}{\zt{AKID}}        %AK's Id
\newcommand{\hmacu}{\zt{HMAC-KU}} %HMAC's uplink key
\newcommand{\hmacd}{\zt{HMAC-KD}} %HMAC's downlink key
\newcommand{\hmac}{\zt{HMAC}}       %HMAC's for general reference
\newcommand{\concat}{\mid}          %Concatenation operator

\newcommand{\tekrequest}{\zt{Request}}   %TEK request message
\newcommand{\tekreply}{\zt{Reply}}       %TEK reply message
\newcommand{\tekreject}{\zt{Reject}}     %TEK reject message

\newcommand{\pkmvrsax}{\ensuremath{P}}
\newcommand{\prepakx}{\ensuremath{c}}
\newcommand{\hmacx}{\ensuremath{d}}
\newcommand{\pkmvsatekx}{\ensuremath{Q}}
\newcommand{\pkmvtekx}{\ensuremath{R}}
\newcommand{\tekx}{\ensuremath{e}}

\newcommand{\labelpmsisynch}{\isynchclaim(\pkmvrsax,\ms)}   % P(ms) isynch
\newcommand{\labelpbsisynch}{\isynchclaim(\pkmvrsax,\bs)}   % P(bs) isynch
\newcommand{\labelpisynch}{\isynchclaim(\pkmvrsax)}         % P isynch
\newcommand{\labelpmssession}{\sessionclaim(\pkmvrsax,\ms,\prepakx)} % P(ms) session
\newcommand{\labelpbssession}{\sessionclaim(\pkmvrsax,\bs,\prepakx)} % P(bs) session
\newcommand{\labelpsession}{\sessionclaim(\pkmvrsax,\prepakx)}       % P session
\newcommand{\labelqmssessiond}{\sessionclaim(\pkmvsatekx,\ms,\hmacx )} % Q(ms) session for d
\newcommand{\labelqbssessiond}{\sessionclaim(\pkmvsatekx,\bs,\hmacx)} % Q(bs) session for d
\newcommand{\labelqsessiond}{\sessionclaim(\pkmvsatekx,\hmacx)} % Q session for d
\newcommand{\labelqmswsessione}{\wsessionclaim(\pkmvsatekx,\ms,\tekx)} % Q(ms) wsession for e
\newcommand{\labelqbswsessione}{\wsessionclaim(\pkmvsatekx,\bs,\tekx)} % Q(bs) wsession for e
\newcommand{\labelqwsessione}{\wsessionclaim(\pkmvsatekx,\tekx)} % Q wsession for e
\newcommand{\labelqmsisynch}{\isynchclaim(\pkmvsatekx,\ms)}   % Q(ms) isynch
\newcommand{\labelqbsisynch}{\isynchclaim(\pkmvsatekx,\bs)}   % Q(bs) isynch
\newcommand{\labelqisynch}{\isynchclaim(\pkmvsatekx)}         % Q isynch
\newcommand{\labelrmsisynch}{\isynchclaim(\pkmvtekx,\ms)}     % R(ms) isynch
\newcommand{\labelrbssynch}{\nisynchclaim(\pkmvtekx,\bs)}      % R(bs) synch
\newcommand{\labelrmssessiond}{\sessionclaim(\pkmvtekx,\ms,\hmacx )} % R(ms) session for d
\newcommand{\labelrbssessiond}{\sessionclaim(\pkmvtekx,\bs,\hmacx)} % R(bs) session for d
\newcommand{\labelrsessiond}{\sessionclaim(\pkmvtekx,\hmacx)} % R session for d
\newcommand{\labelrmswsessione}{\wsessionclaim(\pkmvtekx,\ms,{\tekx'})} % R(ms) wsession for e
\newcommand{\labelrbswsessione}{\wsessionclaim(\pkmvtekx,\bs,{\tekx'})} % R(bs) wsession for e
\newcommand{\labelrwsessione}{\wsessionclaim(\pkmvtekx,{\tekx'})} % R wsession for e
\newcommand{\labelpqisynch}{\isynchclaim(\pkmvrsax\pkmvsatekx)} % PQ isynch
\newcommand{\labelpqsessionc}{\sessionclaim(\pkmvrsax\pkmvsatekx,\prepakx)} % PQ session c
\newcommand{\labelpqwsessione}{\wsessionclaim(\pkmvrsax\pkmvsatekx,\tekx)} % PQ wsession e
\newcommand{\labelpqrisynch}{\isynchclaim(\pkmvrsax\pkmvsatekx\pkmvtekx)} % PQR isynch
\newcommand{\labelpqrsessionc}{\sessionclaim(\pkmvrsax\pkmvsatekx\pkmvtekx,\prepakx)} % PQR session c
\newcommand{\labelpqrwsessione}{\wsessionclaim(\pkmvrsax\pkmvsatekx\pkmvtekx,\tekx)} % PQR wsession e
\newcommand{\labelpqrwsessionep}{\wsessionclaim(\pkmvrsax\pkmvsatekx\pkmvtekx,\tekx')} % PQR wsession e'

\setmsckeyword{}

\newcommand{\mscnslsettings}{
  \setlength{\topheaddist}{1cm}
  \setlength{\instdist}{3cm}
  \setlength{\envinstdist}{1.5cm}
  \setlength{\levelheight}{0.6cm}
  \setlength{\actionwidth}{1.5cm}
}

\newcommand{\mscwimaxsettings}{
  \setlength{\topheaddist}{1.5cm}
  \setlength{\instdist}{9cm}
  \setlength{\envinstdist}{3cm}
  \setlength{\levelheight}{1cm}
  \setlength{\actionwidth}{2cm}
}

\begin{document}
\begin{frontmatter}

\title{A framework for compositional verification of security protocols}

\author[ntnu1]{Suzana Andova\thanksref{ercim}},
\ead{suzana@item.ntnu.no}
\author[ethz]{Cas Cremers}\ead{cremersc@inf.ethz.ch},
\author[ntnu2]{Kristian Gj\o steen\thanksref{nrc}},
\ead{kristian.gjosteen@math.ntnu.no}
\author[unilu]{Sjouke Mauw},
\ead{sjouke.mauw@uni.lu}
\author[ntnu1]{Stig F. Mj\o lsnes},
\ead{sfm@item.ntnu.no}
\author[unilu]{Sa\v{s}a Radomirovi\'{c}\corauthref{cor}\thanksref{ercim}\thanksref{crm}}\ead{sasa.radomirovic@uni.lu}

\corauth[cor]{Corresponding author. Phone: (+352) 46 66 44 5484, Fax: (+352) 46 66 44 5500}
\thanks[ercim]{This work was partially carried out during the
tenure of an ERCIM Fellowship.}
\thanks[nrc]{Supported in part by
the Norwegian Research Council project 158597 NTNU Research
Programme in Information Security.}
\thanks[crm]{Supported in part by
a Centre de Recerca Matem\`atica Postdoctoral Fellowship.}

\address[ntnu1]{
    Dept.~of Telematics,
    NTNU,
    N-7491 Trondheim,
    Norway.}
\address[ntnu2]{
    Dept.~of Mathematical Sciences,
    NTNU,
    N-7491 Trondheim,
    Norway.}
\address[ethz]{
    Dept.~of Computer Science,
    ETH Z\"urich,
    8092 Z\"urich,
    Switzerland.}
\address[unilu]{
    Universit\'e du Luxembourg,
    Facult\'e des Sciences, de la Technologie et de la Communication,
    6, rue Richard Coudenhove-Kalergi,
    L-1359 Luxembourg.}

\begin{abstract}
Automatic security protocol analysis is currently feasible only for small
protocols. Since larger protocols quite often are composed of many small
protocols, compositional analysis is an attractive, but non-trivial
approach.

We have developed a framework for compositional analysis of a large
class of security protocols. The framework is intended to facilitate
automatic as well as manual verification of large structured security
protocols.
Our approach is to verify properties of component protocols in a
multi-protocol environment, then deduce properties about the composed
protocol. To reduce the complexity of multi-protocol
verification, we introduce a notion of protocol independence and prove
a number of theorems that enable analysis of independent component protocols in
isolation.

To illustrate the applicability of our framework to real-world protocols, we study a key establishment sequence in \wimax
consisting of three subprotocols. Except for a small amount of trivial
reasoning, the analysis is done using automatic tools.

\end{abstract}

\end{frontmatter}

\section{Introduction}
\label{sec:introduction}

Security protocols are a crucial component of many contemporary
applications. Their security is however very difficult to assess for
humans, mainly due to the vast number of attack options available to
an adversary. To deal with this complexity, a structured approach is
needed. Starting from abstract protocols, formal methods faciliate the
systematic detection of attacks or the generation of a proof of
correctness. Automating this process in order to minimize the risk of
human error is one of the major goals in security protocol analysis.

Automatic protocol verification is, in general, a complex task even
for short protocols. The time needed for verification of a protocol
using modern methods employed by state of the art tools such as
Scyther~\cite{wwwscyther} or AVISPA~\cite{avispatool} is still
exponential with respect to the number of messages.  Consequently,
automatic verification of large protocols is currently infeasible.
In this paper, we attempt to narrow the gap between small,
academic protocols and large, industrial protocols by taking advantage of
compositional verification.

Large protocols are usually built from structured components.
They typically consist of several
(optional) protocols composed in parallel, or a sequential composition of a key
establishment protocol and a secure data transfer protocol that uses
the key. For instance, IPSec, SET, and \wimax have all been designed
with such a principle in mind. A compositional approach to the
design and analysis of security protocols is therefore natural and 
expected to reduce the complexity of the analysis of the large
protocol to the order of the complexity of the analysis of the
largest component.
This could be achieved by first verifying properties of the components
in isolation and then using the results to deduce properties of the composed
protocol. 
However, as no generic compositionality results are known, further
assumptions are needed to facilitate this type of 
reasoning.
\note{CC}{Rephrased. (Unless you want this paper to be forever cited (by
papers that do fix it) as the 'they thought it couldn't be done' paper.}
\note{CC}{I'm not happy with SR's version but mine also is clunky. Maybe
we should revert this.}
\note{SR}{I am happy with yours and would suggest adding ``naive'' in front of approach for mine. The only thing I want to achieve is a connection between this and the following paragraph.}

\mscnslsettings
\begin{figure}[htb]
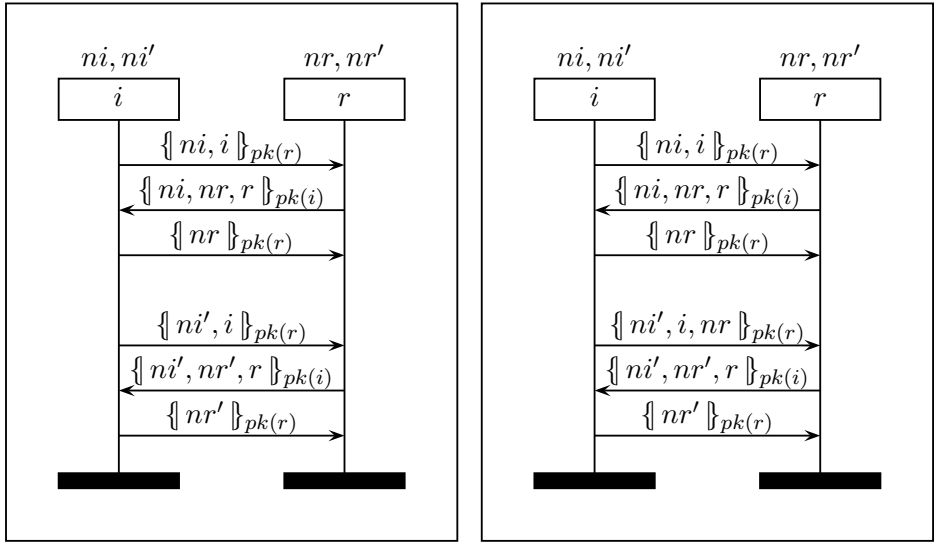

\centering{ {\small
\begin{msc}{}

\declinst{i}{$\noni,\nonip$}{$\irole$}
\declinst{r}{$\nonr,\nonrp$}{$\rrole$}

\mess{$ \enc{\noni,\irole}{\pkr} $}{i}{r} \nextlevel \mess{$
\enc{\noni,\nonr,\rrole}{\pki} $}{r}{i} \nextlevel \mess{$
\enc{\nonr}{\pkr} $}{i}{r} \nextlevel[2]

\mess{$ \enc{\nonip,\irole}{\pkr} $}{i}{r} \nextlevel \mess{$
\enc{\nonip,\nonrp,\rrole}{\pki} $}{r}{i} \nextlevel \mess{$
\enc{\nonrp}{\pkr} $}{i}{r}

\end{msc}
~%
\begin{msc}{}

\declinst{i}{$\noni,\nonip$}{$\irole$}
\declinst{r}{$\nonr,\nonrp$}{$\rrole$}

\mess{$ \enc{\noni,\irole}{\pkr} $}{i}{r} \nextlevel \mess{$
\enc{\noni,\nonr,\rrole}{\pki} $}{r}{i} \nextlevel \mess{$
\enc{\nonr}{\pkr} $}{i}{r} \nextlevel[2]

\mess{$ \enc{\nonip,\irole,\nonr}{\pkr} $}{i}{r} \nextlevel
\mess{$ \enc{\nonip,\nonrp,\rrole}{\pki} $}{r}{i} \nextlevel
\mess{$ \enc{\nonrp}{\pkr} $}{i}{r}

\end{msc}
}} \caption{Repeated \nsl{} protocol: incorrect and correct
chaining.} \label{fig:nslsquare}
\end{figure}

\begin{figure}[htb]
\centering{ {\small
\psset{xunit=.3cm,yunit=.3cm}
\begin{pspicture}(0,0)(25,19)
\psframe(0,0)(25,19)
\psset{linewidth=0.6pt}
\psline(3,6)(3,14)   %
  \rput(3,15.5){$A(i)$}
  \psframe(2,14)(4,14.5)  %
  \psframe*(2,5.7)(4,6)   %
  \psline{->}(3,13)(5,13)
  \psline{<-}(3,12)(5,12)
  \psline{->}(3,11)(5,11)
  \psline{->}(3,9)(5,9)
  \psline{<-}(3,8)(5,8)
  \psline{->}(3,7)(5,7)
\psline(14,1)(14,9)  %
  \rput(14,10.5){$B(r)\sharp 1$}
  \psframe(13,9)(15,9.5)  %
  \psframe*(13,0.7)(15,1)   %
  \psline{->}(12,8)(14,8)
  \psline{<-}(12,7)(14,7)
  \psline{->}(12,6)(14,6)
  \psline{->}(12,4)(14,4)
  \psline{<-}(12,3)(14,3)
  \psline{->}(12,2)(14,2)
\psline(22,8)(22,16) %
  \rput(22,17.5){$B(r)\sharp 2$}
  \psframe(21,16)(23,16.5)  %
  \psframe*(21,7.7)(23,8)   %
  \psline{->}(20,15)(22,15)
  \psline{<-}(20,14)(22,14)
  \psline{->}(20,13)(22,13)
  \psline{->}(20,11)(22,11)
  \psline{<-}(20,10)(22,10)
  \psline{->}(20,9)(22,9)
\psset{linewidth=0.4pt,linecolor=gray,linearc=5}
  \psline(5,13)(7,13)(18,15)(20,15) %
  \psline(5,12)(7,12)(18,14)(20,14)
  \psline(5,11)(7,11)(18,13)(20,13)
\psset{linearc=3}
  \psline(5,9)(7,9)(10,8)(12,8) %
  \psline(5,8)(7,8)(10,7)(12,7)
  \psline(5,7)(7,7)(10,6)(12,6)
\psset{linewidth=0.6pt,linestyle=dotted,linecolor=gray,linearc=2}
  \psline(18,11)(20,11) %
  \psline(18,10)(20,10)
  \psline(18,9)(20,9)
  \psline(10,4)(12,4) %
  \psline(10,3)(12,3)
  \psline(10,2)(12,2)
\end{pspicture}
}} \caption{Authentication problem in incorrectly chained NSL protocol.}\label{fig:nslsquareerror}
\end{figure}
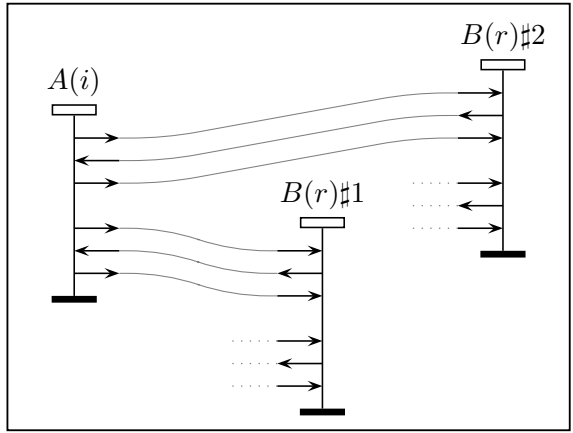

We illustrate the non-triviality of protocol composition by means of the
well-known Needham-Schroeder-Lowe (\nsl) public key authentication
protocol~\cite{needham78using,lowe96a}. In isolation, it satisfies
even the strongest forms of authentication, such as agreement and
synchronization~\cite{CrMaVi06}. However, when sequentially
composing this protocol with itself (see the left drawing in
Figure~\ref{fig:nslsquare}), authentication is not preserved. The
reason is that the initiator $\irole$ may successfully finish his
run of the composed protocol, while the responder $\rrole$ possibly
never executed the second half of the protocol. This is because the
second half of the initiator's run may match to the first half of a
different run of the responder. This authentication problem is
illustrated in Figure~\ref{fig:nslsquareerror}. Here we see agent
$A$ executing the initiator role $i$ and agent $B$ executing two
different runs of the responder role $r$. The intruder links the
messages as indicated. Run $A(i)$ and run $B(r)\sharp 2$ will agree
on the values of $\noni$, and $\nonr$, but not on the values of
$\nonip$ and $\nonrp$, since these last two values are not
communicated between these two runs. In a similar way, it is clear
that run $A(i)$ and run $B(r)\sharp 1$ do not agree on the
supposedly shared nonces.

This problem is solved in the right drawing in
Figure~\ref{fig:nslsquare} by chaining the two protocols. A nonce
from the first instance of \nsl{} is repeated as payload in the
second instance. In this way the two protocols become linked and
the chained protocol satisfies authentication.
The authentication problem from Figure~\ref{fig:nslsquareerror} is now
impossible.

Even though it is well known that the composition of secure protocols
is in general not
secure~\cite{heintze96comp,kelsey97protocol,Alves-Foss99a,Cr06} and
compositionality has been recognised as one of the open challenges for
security protocol analysis~\cite{meadows01open,Cr04a}, the vast
majority of formalisms and tools for security protocols have only
addressed single-protocol (i.e.~non-composed) analysis and
verification. Early work on identifying and addressing the problem
includes \cite{DBLP:journals/jcs/Paulson98,DBLP:conf/csfw/DurginMP01}.
An initial attempt within the Strand Spaces
model~\cite{thayer99mixed} has led to some theoretical results
about compositionality. The Strand Spaces approach is similar to the one
taken here in that both attempt to identify the abstract properties two
protocols need to satisfy in order to be securely composable. However,
this work significantly improves upon the Strand Spaces approach in
terms of efficiency in verifying composed protocols and by considering
sequential composition, which was absent in the Strand Spaces model. 
One of the recent significant developments in
compositional protocol analysis is Protocol
Composition Logic (PCL) \cite{DDMR07,mitchell03comp}. %
It provides support for compositional reasoning, and has been applied in a number of case studies, including
the verification of the TLS and IEEE 802.11i protocols
\cite{DBLP:conf/ccs/HeSDDM05} and contract signing protocols
\cite{DBLP:journals/tcs/BackesDDMT06}. While the PCL approach is quite general, it cannot, in contrast to the present approach, be easily automated.

In this paper, we develop a framework to verify security
properties of protocols that are composed from several smaller
protocols. %
We prove several theorems concerning the deduction of properties
of a sequential composition of two protocols from properties
these protocols have when
running together in a multi-protocol environment. With these theorems,
we reduce the analysis of a sequential composition to the analysis
of the component protocols running together.

Analysing several protocols in a multi-protocol environment is, in
general, no
easier than analysing their sequential composition. %
 In order to make
automatic analysis feasible, we introduce the notion of protocol set
independence, where ciphertexts, signatures, and message authentication
tags originating in one protocol set will never be accepted by the
other protocol set and vice versa. This notion allows us to prove
several theorems regarding the deduction of properties of protocols
running together in a multi-protocol environment from properties these
protocols have when running in isolation.

Verifying independence itself is non-trivial,
therefore we need the notion of strong independence, where the
forms of ciphertexts, message authentication tags, and
signatures in the two protocol sets are sufficiently different to
prevent confusion. Strong independence can be easily verified at the
syntactical level, and implies independence. We show that through
common design strategies for security protocols in current use,
strong independence will be satisfied. Note that different protocols
can use the same cryptographic keys and still be both, independent
and strongly independent.

The model we use is
based on the operational semantics for security protocols defined
in~\cite{CrMa04b}. In contrast to other approaches, in which only
singular protocols are considered, this model provides a semantics
of protocols in a multi-protocol setting.
This makes it a good
starting point for compositional verification, since, as indicated,
the problem of
proving correctness of a composed protocol can be
translated into the problem of proving correctness of the components in a
multi-protocol setting comprising the components themselves.

To show the applicability of our work, we perform a case study. We
have chosen to focus on the IEEE~802.16 standard, also known as
\wimax. This standard specifies the air interface of wireless access
systems featuring a security sublayer intended to protect network
operators from theft of service and provide confidentiality to
subscribers. \wimax features a security sublayer consisting of
several subprotocols for authentication, key management, and secure
communication. This makes \wimax well suited for an analysis in our
framework. Our verification is completely tool-supported, except for
some trivial reasoning and theorem application. %

\subsection*{Overview of the paper}

We start off by giving a brief description of the security
protocol model and security properties used in
Section~\ref{sec:framework}.  In Section~\ref{sec:reasoning}, we
develop a framework for compositional reasoning about security
protocols, and prove a number of compositionality theorems. We
show how the developed theory can be applied in practice by
performing a case study on key management protocols in the
security sublayer of \wimax in Section~\ref{sec:wimax}. Related
work is discussed in Section~\ref{sec:relatedwork}, and we draw
conclusions and discuss future work in Section~\ref{sec:conclusion}.

\section{Security Protocols and Their Semantics}
\label{sec:framework}

In this section we describe an existing formal framework for modeling
security protocols, and extend it with notions relevant for
compositional reasoning.

We begin by giving a brief overview of the model in
Section~\ref{sec:overview} before describing the full technical details
in Sections~\ref{secprot} and~\ref{runs}. The model presented here is based on the
model defined in~\cite{CrMa04b}. Readers who are familiar with the basic
model may skip to Section~\ref{after-the-model} on
page~\pageref{after-the-model}, as the only change is the introduction of
parameters for protocols.

In Sections~\ref{sec-trace-restriction} and~\ref{security-properties} we
further extend the model with features not present in the basic
model defined in~\cite{CrMa04b}, namely trace restrictions (similar to
preconditions in PCL and elsewhere), satisfiability predicates,
and new security notions.

\subsection{Overview}
\label{sec:overview}

The basic entities in our framework are {\em role specifications}.
Every role specification consists
of a sequence of uniquely \emph{labeled} \emph{events} describing
the messages an agent shall send and receive, when it executes 
the role specification, as well as certain
\emph{security claims}. The role specification includes
\emph{constants} which roughly correspond to nonces,
\emph{variables} which store values read from the network, and
\emph{parameters} which represent input.

A \emph{protocol} is a collection of role specifications that
communicate by sending and receiving messages. More precisely,
a protocol is a partial function, mapping role names to role
specifications. A \emph{run} is an execution of a role
specification by an \emph{agent}. Communication between runs is
asynchronous and is modeled by agents reading messages from and
writing messages to a shared input/output buffer (by executing
read and send events). As the buffer is completely under the
control of the adversary, according to the standard Dolev-Yao
intruder model, we identify the buffer with the intruder
knowledge. The actual behavior of the entire system, consisting of
the intruder and a set of agents executing a number of runs, is
encoded in the \emph{traces} of the system. In some situations, we
are not interested in all possible traces but in a subset of
traces that have a certain property; for instance, the subset of
traces whose input values are secret. In that case, we talk about
\emph{trace restriction}.

Security properties in our framework are local to a role and are
described by the \emph{claim} events in the role specifications.
Every claim event in a trace results in a statement about the
trace that may or may not be true. In this paper, we focus on
three security properties: \emph{secrecy}, \emph{authentication},
and \emph{session key establishment}. A secrecy claim event is essentially
the statement that something never enters the adversary's
knowledge, as determined by the trace. Authentication is captured
by the notion of \emph{synchronization}. A synchronization claim
event translates into the statement that there are runs for the
other protocol roles in the trace with read and send events that
match this run's send and read events exactly, both in content and
in order.
Our notion of \emph{session keys} is that a session key is secret and
identifies a protocol session, in the sense that there is exactly one
execution of every protocol role sharing the session key.

%
%
%
%

\iffalse We continue the section with a brief description of the
model for security protocols as defined in~\cite{CrMa04b}. In
Section~\ref{secprot} we define what security protocol in our
model means, and in Section~\ref{runs} we define the trace based
operational semantics for security protocols.
Section~\ref{security-properties} and \ref{sec-trace-restriction}
extend the model of~\cite{CrMa04b}. In
Section~\ref{security-properties} the notion of a security
property is given together with a number of specific security
properties used throughout the paper. Finally, in
Section~\ref{sec-trace-restriction} we discuss the necessity of
allowing protocols to accept inputs, often required to be of
specific type, which we represent via trace restrictions. \fi

\subsection{Security Protocol Specification}
\label{secprot}

Let $\idset$ be a set of \emph{identifiers}, $\roleset$ a set of
\emph{role names} or \emph{roles} for short, and $\funcset$ a set
of (global) functions. There are three types of identifiers:
\emph{constants}, \emph{variables}, and \emph{parameters}.
Constants include the general notion of nonces, and we will
informally refer to some constants as nonces.
Concatenation or tupling of terms is written as $(x,y)$.
Encryptions of a term $x$ with a term $y$ are denoted by $\enc{x}{y}$.
\emph{Role terms}
can be considered as templates for messages that are read or sent
by the agents. The set of role terms is defined as:
\begin{multline*}
\roletermset ::=  \idset \mid \roleset \mid \funcset(
\roletermset^*) \\ \mid (\roletermset, \roletermset) \mid
\enc{\roletermset}{\roletermset}
\end{multline*}
Terms that have been
encrypted with a term, can only be decrypted by its inverse term
which is either the same term (for symmetric encryption) or the
inverse key (for asymmetric encryption). We use $k\inv$ to denote
the inverse key of a key $k$. In this work, functions from $\funcset$ are only used
to construct long-term keys, such as $\pkr$, $\ski$, $k(x,y)$.
Short term session keys are represented by constants. In the
remainder of the paper $x, y , z$ range over $\roletermset$, and
$c, d$ over the $\idset$ set.

\begin{exmp}
The first message sent by the initiator in the \nsl{}
protocol is denoted by $\enc{\noni,\irole}{\pkr}$, where
$\noni \in \idset$ is a constant, $\irole,\rrole \in \roleset$ are
role names, and $\publickey{} \in \funcset$.
\end{exmp}

We say that $x_1$ is a \emph{subterm} of $x_2$ if $x_1 \subterm
x_2$, where $\subterm$ is the smallest transitive relation
satisfying the following rules, for all terms $x_1,x_2$:
\begin{equation*}
  x_1 \subterm x_1, \quad
  x_1 \subterm (x_1,x_2), \quad
  x_2 \subterm (x_1,x_2), \quad
  x_1 \subterm \enc{x_1}{x_2}, \quad
  x_2 \subterm \enc{x_1}{x_2} \text.
\end{equation*}

For a given set of \emph{labels} $\labelset$ and a set of
\emph{claims} $\claimset$ we define the set of \emph{events}
$\eventset$ as:
\begin{multline*}
\eventset  = \bigl\{
  \createevent_{\thelabel} (r),
  \sendevent_{\thelabel} (r,r',x), \readevent_{\thelabel} (r',r,x),
  \claimevent_{\thelabel} (r,c \,[,x]) ,
  \stopevent_{\thelabel} (r)
  \bigm| \\
 {\thelabel} \in \labelset, r,r' \in \roleset, x \in \roletermset, c \in \claimset
\bigr\}
\end{multline*}
The labels $\thelabel$ extending the events are needed to
disambiguate multiple occurrences of an event in protocol
specifications (e.g.\ when composing two instances of the NSL
protocol). A second use of these labels is to express which send and
read events are supposed to correspond (e.g.\ in NSL the first message
sent by the initiator is linked to the first message received by the
responder).

Event $\sendevent_{\thelabel} (r,r',x)$ denotes the sending of
message $x$ by $r$, apparently to $r'$. Likewise,
$\readevent_{\thelabel} (r',r,x)$ denotes the reception of message
$x$ by $r'$, apparently sent by $r$. We interpret role terms of
the form $\enc{u}{v}$ in a $\sendevent$ event as encryption with
symmetric or public encryption keys, or signing with private
signing keys. In a $\readevent$ we interpret it as decryption with
symmetric or private encryption keys, or verification with public
signing keys. An agent can encrypt or decrypt a term only when it
has the relevant key in its knowledge. Event
$\claimevent_{\thelabel} (r,c \,[,x])$ expresses that $r$ upon
execution of this event expects the security property associated
with the claim $c$ to hold with optional argument $x$. A claim event
is always local to a role, and does not imply that other roles
expect the security property associated with the claim $c$ to hold
for them.
Events $\createevent_{\thelabel}(r)$ and $\stopevent_{\thelabel}(r)$
are used to signal the start and end of the role.

\begin{exmp}
The first send event of the initiator in the \nsl{} protocol is denoted by
$\sendevent_{\thelabel_1} (\irole, \rrole, \enc{\noni,\irole}{\pkr})$, where
$\noni \in \idset$ is a constant and $\thelabel_1$ is some label.
The first read event of the responder in the NSL protocol is denoted by
$\readevent_{\thelabel_1} (\irole, \rrole, \enc{\noni,\irole}{\pkr})$, where
$\noni \in \idset$ is a variable.
\end{exmp}

A \emph{role specification} is a pair $(\elist, \idtyper)$ where
$\elist\in\eventset^*$ is a list of events and $\idtyper: \idset
\rightarrow \{ \consttype, \paramtype, \vartype \}$ is a function
that assigns types to the identifiers that appear in $\elist$. We
require that there is only one $\createevent$ and one $\stopevent$
in the event list, and that they start and terminate the list.
Furthermore, we require that the role names in the $\createevent$
and $\stopevent$ events are the same, and they match the role
names that appear in $\claimevent$ events and  the sender and
recipient, respectively, in $\sendevent$ and $\readevent$ events.
This is the specification's \emph{role name}. The set of all role
specifications is denoted by $\rolespec$.

Note that only in the context of a role specification $\rs$ can we
talk about the set of variables or parameters. For a role
specification $\rs = (\elist, \idtyper)$, we write
$\varterm{\rs}{x}$ for the set of identifiers that appear in role
term $x$ and are considered variables in the
role specification.%

A \emph{protocol} is a partial mapping of role names to role
specifications, i.e.~$\roleset \rightarrow \rolespec$. We say that
$r$ is a role in protocol $P$ if $r\in \domain(P)$, the domain of $P$.
If $r$ is a role in protocol $P$ and $\thelabel$ is a label of an event in the
event list of $r$ then we write $\thelabel \in P(r)$. We extend
this notation in the obvious way to $\thelabel \in P$. By
$\idset(P)$ we denote the set of all identifiers that appear in
protocol $P$. The universe of protocols is denoted by $\protocols$.

For a protocol $P$, we require that all labels are unique, except for the labels of corresponding read and send events which have to be identical.
 For a set of protocols $\Pi$, we
require that a label is used in at most one protocol.

We define a relation $\before'$ on the events of a protocol as the
union of the obvious event orders on the role specifications. We
extend this relation with all pairs of identically labeled
$\sendevent$ and $\readevent$ events so that such $\sendevent$
events always precede the corresponding $\readevent$ events.
The partial order $\before$ is the transitive
closure of $\before'$ and represents causality preorder.

\begin{exmp}
\label{ex:nslprime} The following example specifies the \nsl'{}
protocol, which is the bottom right subprotocol in
Figure~\ref{fig:nslsquare}. Notice the parameter $\nonr$ and the
fact that $\nonip$ is considered a constant by role $\irole$,
whereas it is a variable for role $\rrole$.
\[
  \begin{array}{lll}
    \nsl'(\irole) & = &
      (
      \createevent_{1} (\irole) \seq
      \sendevent_{2} (\irole, \rrole, \enc{\nonip,\irole,\nonr}{\pkr}) \seq \\
      && \readevent_{3} (\rrole, \irole, \enc{\nonip,\nonrp,\rrole}{\pki}) \seq
      \sendevent_{4} (\irole, \rrole, \enc{\nonrp}{\pkr}) \seq
      \stopevent_{5} (\irole)
      ,\\
      & &
      \{ \nonr \mapsto \paramtype, \nonip \mapsto \consttype,
      \nonrp \mapsto \vartype \}
      )
      \\
    \nsl'(\rrole) & = &
      (
      \createevent_{6} (\rrole) \seq
      \readevent_{2} (\irole, \rrole, \enc{\nonip,\irole,\nonr}{\pkr})
      \seq \\
      && \sendevent_{3} (\rrole, \irole, \enc{\nonip,\nonrp,\rrole}{\pki}) \seq
      \readevent_{4} (\irole, \rrole, \enc{\nonrp}{\pkr}) \seq
      \stopevent_{7} (\rrole)
      ,\\
      & &
      \{ \nonr \mapsto \paramtype, \nonip \mapsto \vartype,
      \nonrp \mapsto \consttype \}
      )
      \\
  \end{array}
\]
\end{exmp}

\subsection{Runs and Traces}
\label{runs}

In this section we describe how, through instantiation, an
abstract role specification can be transformed into an execution
of a role, which we call a \emph{run}. Furthermore, we define how
the interleaved operation of a collection of runs defines the
\emph{traces} of a system.

\emph{Run terms} model the actual messages sent in a protocol.
Since the run terms are instantiations of role terms they are
defined similarly. Let $\runidset$ be a set of \emph{run
identifiers}, $\IT$ a set of \emph{intruder-generated} run terms
and $\agentset$ a set of \emph{agent names} which is a disjoint
union of a set of trusted and a set of untrusted agents,
$\trustedagentset$ and $\untrustedagentset$ respectively. The set
of run terms is defined as:
\begin{multline*}
\runtermset ::= \agentset \mid \funcset(\runtermset^*) \mid
\idset\markid{\runidset} \mid \IT \mid \\ %
(\runtermset, \runtermset) \mid \enc{\runtermset}{\runtermset}
\end{multline*}
The run terms
of the form $\idset\markid{\runidset}$ and the terms in $\IT$ are
called \emph{nonce run terms}.
The subterm relation $\subterm$ on
run terms is defined similarly to the subterm relation on role
terms. Since it is clear from the context which one is used, we
allow the same notation for both relations. In the remainder of
the paper $t, u , v$ range over $\runtermset$.

A role term is turned into a run term when abstract role names are
replaced by concrete agent names, and constants are made unique by
extending them with a run identifier. This is done by means of an
\emph{instantiation}, which is a triplet $(rid, \rho, \sigma)$,
where $rid \in \runidset$, $\rho$ is a partial function from role
names to agent names, and $\sigma$ is a partial function from
identifiers to run terms. We denote the set of
all possible instantiations by $\Inst$.

In the context of some role specification $\rs$ with type function
$\idtyper$, an instantiation $\z{inst}=(rid, \rho, \sigma)$ turns
a role term $x$ into a run term, if $\rho$ is defined for every
role name that appears in $x$ and $\varterm{\rs}{x} \subseteq
\domain(\sigma)$. For any $f \in \funcset$ and role terms
$x_1,\ldots,x_n \in \roletermset$, instantiation is defined
recursively by:
\begin{equation*}
  \z{inst}(x)  = \left\{
      \begin{array}{ll}
     \rho(r)        & \mbox{if } x\equiv r \in \roleset\\
     c\markid{rid}      & \mbox{if } x \equiv c \in \idset \land
     \idtyper(c) = \consttype \\
     \sigma(x)      & \mbox{if } x \in \idset \land \idtyper(x) \in \{
     \paramtype, \vartype \}\\
     f(\z{inst}(x_1), \ldots ,
       \z{inst}(x_n) )
                            & \mbox{if } x \equiv f(x_1, \ldots, x_n)
                \\
     (\z{inst}(x_1),\z{inst}(x_2))
                & \mbox{if } x \equiv (x_1,x_2) \\
     \enc{\z{inst}(x_1)}{\z{inst}(x_2)}
                & \mbox{if } x \equiv \enc{x_1}{x_2} \\
      \end{array}
  \right.
\end{equation*}
If an instantiation cannot be applied because $\varterm{\rs}{x}
\not\subseteq \domain(\sigma)$, we say that $x$ has \emph{free
  variables} in this context.

\begin{exmp}
If we apply instantiation
$(42, \{ \irole \mapsto a, \rrole \mapsto b \},
  \{ \noni \mapsto \noni\markid{41} \}
)$
to the contents of the first send event of the responder in the \nsl{} protocol
$\enc{\noni,\nonr,\rrole}{\pki}$,
we obtain
$\enc{\noni\markid{41},\nonr\markid{42},b}{\publickey{a}}$,
\end{exmp}

Instantiations are essential ingredients to define the notion of a
\emph{run} of a role. A \emph{run} of a role specification $\rs =
(\elist,\idtyper)$ is a pair $(inst, \elist')$, where $inst \in
\Inst$ and $\elist'$ is a suffix of $\elist$. In this definition
we express that one is mainly interested in the current state of
an agent executing a role. We model this dynamic aspect by
requiring that the list $\elist' \in \eventset^*$ contains the
remaining events in the role specification, and not the complete
role specification. The instantiation $inst$ contains the actual
values of the variables and parameters, as well as the agent names
expected to execute the other protocol roles. The set of all runs
is denoted by $\run$. A \emph{run event} is a pair
$(inst,ev)\in\Inst\times \eventset$. These are the events that can
be observed when executing a system. A system's behavior is
represented by a sequence of run events, which we call a
\emph{trace}. The universe of traces is denoted by $\traces$.

Let $P$ be a protocol with a role specification $\rs =
(\elist,\idtyper)$, and let $inst=(rid, \rho, \sigma)$ be an
instantiation. The pair $(inst, \elist)$ is an \emph{initial run}
for $\rs$ if and only if $\domain(\rho) = \domain(P)$ and
$\domain(\sigma)=\idtyper^{-1}(\paramtype)$ ($\sigma$ is defined
for all role parameters, specifying the run's input). The set of
all initial runs for all roles of a protocol $P$ is denoted by
$\runsof(P)$. For a protocol set $\Pi$, we let
\begin{equation*}
  \runsof(\Pi) = \bigcup_{P \in \Pi} \runsof(P) \text.
\end{equation*}

\begin{exmp}
An initial run of the initiator of the \nsl'{} protocol from
Example~\ref{ex:nslprime} is
$(
    (42, \{ \irole \mapsto a, \rrole \mapsto b \},
    \{ \nonr \mapsto \noni\markid{41}, \nonrp \mapsto \undef \})
 ,
    \createevent_{1} (\irole) \seq$ \\
    $\sendevent_{2} (\irole, \rrole, \enc{\nonip,\irole,\nonr}{\pkr}) \seq$
    $\readevent_{3} (\rrole, \irole, \enc{\nonip,\nonrp,\rrole}{\pki}) \seq$
    $\sendevent_{4} (\irole, \rrole, \enc{\nonrp}{\pkr}) \seq$
    $\stopevent_{5} (\irole)
)$.
Notice that $\nonr$ is the only parameter (and thus must be
initialized), that variable $\nonrp$ has no initial value and that
$\nonip$ is a constant.
\end{exmp}

For a protocol set $\Pi$ we consider a system with a number of runs
(communicating with each other) executed by agents in presence of an
intruder. We assume a standard Dolev-Yao model, in which the
intruder has complete control over the communication network. The
knowledge of the intruder, denoted by $\IK$, is a subset of run
terms. He can decrypt messages if he knows the appropriate
decryption key, and he can construct messages from his knowledge
set. We express this by requiring that $\IK$ is closed, that is:
\begin{align*}
\forall_{u,v \in \IK} (u,v) \in \IK & \Rightarrow \enc{u}{v} \in \IK \\
\forall_{u,v} \enc{u}{v}, v\inv \in \IK & \Rightarrow u \in \IK \\
\forall_{u,v} (u,v) \in \IK & \Leftrightarrow u,v \in \IK
\end{align*}
The closure $\closure{\IK}$ of a set of run terms $\IK$ is the
smallest closed superset of $\IK$.

Due to the dynamic behavior of the system, the intruder knowledge
increases during the execution.
We assume that the initial knowledge $\IK_0$ of the
intruder can be derived from the protocol and the context (e.g.
the public keys of all agents and the secret keys of all
compromised agents). We require that $\IT \subseteq \IK_0$. The
derivation of the initial intruder knowledge from the protocol
specification is treated in detail in~\cite{Cr06b}.

The behavior of the system is defined as a transition relation
between system states. Every state is determined by an intruder
knowledge set $\IK$ containing run terms (which is also used to
model an asynchronous communication between agents), and a set $F$
containing all active runs. We denote by $\runids(F)$ the set of
all run identifiers that appear in $F$. Every transition is
labeled with a run event $(inst,ev)\in\Inst\times \eventset$.

The derivation rules for the system are given in
Table~\ref{sosrules}. We denote by $F[x/y]$ the set obtained from
$F$ when $x$ replaces $y$. Note that from run events used to label
transitions, one can uniquely determine role specifications. All
instantiations that appear in a rule are applied in the context of
this role specification.

The $\mathit{create}$ rule expresses that a new run can only be
created if its run identifier has not been used yet. The
$\mathit{end}$ and $\mathit{claim}$ rules express that these
events can always be executed.  Recall that $\closure{\IK}$
denotes the closure of the set $\IK$. The $\mathit{send}$ rule
states that if a run executes a send event, the sent message
(obtained by instantiating a role term in the role specification
context determined by the run event) is added to the intruder
knowledge and the executing run proceeds to the next event.

The $\mathit{read}$ rule determines when a read event can be
executed, with the help of a match predicate defined as follows:
\begin{align*}
  \Match(inst,m,t,inst') \equivalent &
  inst = (rid,\rho,\sigma) \wedge inst' = (rid,\rho, \sigma') \wedge
  \sigma \subseteq \sigma' \wedge {} \\
  & \domain(\sigma') = \domain(\sigma)
  \cup \varterm{\rs}{m} \wedge inst'(m) = t \text.
\end{align*}
The match
predicate decides if an incoming message $t$ can be matched
against a pattern specified by a role term $m$. With respect to
the first instantiation $\z{inst}$, the pattern may contain free
variables. The idea is that the second instantiation $\z{inst}'$
extends the first instantiation by assigning values to the free
variables such that the incoming message equals the instantiated
role term. Note that the run event determines role specification
$\rs$.

\begin{exmp}
We have $\Match(inst,m,t,inst')$ for
$inst =
  (42, \{ \irole \mapsto a, \rrole \mapsto b \},
  \{ \nonr \mapsto \undef \})$,
$m =
  \enc{\noni,\nonr,\rrole}{\pki}$,
$t =
  \enc{\noni\markid{42},\nonr\markid{12},b}{\publickey{a}}$,
and
$inst' =
  (42, \{ \irole \mapsto a, \rrole \mapsto b \},
  \{ \nonr \mapsto \nonr\markid{12} \})$.
This models the first receive event of agent $a$ executing the
initiator role of \nsl{} in run 42. The symbol $\undef$ means that no
value is assigned.
\end{exmp}

A state transition is the conclusion of an application
of one of these rules. In this way, starting from the initial state
$\Sigma_0 = \langle \IK_0, \emptyset \rangle$, where $\IK_0$
refers to the initial intruder knowledge, we can derive all
possible behaviors of a system executing a protocol set  $\Pi$.

\begin{table}[!htb]
\rule{\linewidth}{1pt}
$$
\sosrule
  {create}
  {
    run = (inst, \createevent_{\thelabel}(r) \cdot \elist) \in \runsof(\Pi),
    inst = (rid, \rho, \sigma),
    rid \not\in \runids(F)
  }
  {\transition
    {\langle \IK,F\rangle}
    {(inst, \createevent_{\thelabel}(r))}
    {\langle \IK, F \cup \{(inst, \elist)\}\rangle}}
$$
\vspace{0.1ex}

$$
\sosrule
  {\stopevent}
  {
    run = (inst, \stopevent(r) ) \in F,
  }
  {\transition
    {\langle \IK,F\rangle}
    {(inst, \stopevent_{\thelabel}(r))}
    {\langle \IK, F[(inst,\noevent)/run] \rangle}}
$$
\vspace{0.1ex}

$$
\sosrule
  {send}
  {run = (inst, \sendevent_{\thelabel}(m) \cdot \elist) \in F }
  {\transition
    {\langle \IK,F\rangle}
    {(inst, \sendevent_{\thelabel}(m))}
    {\langle \closure{\IK \cup \{inst(m)\}},
       F[(inst, \elist)/run] \rangle}}
$$
\vspace{0.1ex}

$$
\sosrule
  {read}
  {run = (inst, \readevent_{\thelabel}(m) \cdot \elist) \in F,
    t \in \IK,
    \Match(inst, m,t,inst')}
  {\transition
    {\langle \IK,F\rangle}
    {(inst',\readevent_{\thelabel}(m))}
    {\langle \IK,
       F[(inst', \elist)/run] \rangle}}
$$
\vspace{0.1ex}

$$
\sosrule
  {claim}
  {run = (inst, \claimevent_{\thelabel}(r,c \,[,x]) \cdot \elist) \in F}
  {\transition
    {\langle \IK,F\rangle}
    {(inst,\claimevent_{\thelabel}(r,c \,[,x]))}
    {\langle \IK,F[(inst, \elist)/run]\rangle}}
$$
\vspace{0.1ex}

\rule{\linewidth}{1pt} \caption{Derivation rules.}
\label{sosrules}
\end{table}

We define the \emph{set of traces} generated by the above
derivation rules as a subset of $\traces$.
Let $\alpha \in \traces$ be a trace
of length $\length{\alpha}=n$, and denote by $\alpha_i$ the $i$th
run event in $\alpha$ (starting with $0$).  Then
$\alpha$ is a valid trace for the system if
there exist states $\Sigma_1,\Sigma_2,\dots,\Sigma_n$ such that
$\Sigma_0 \stackrel{\alpha_0}{\rightarrow} \Sigma_1
\stackrel{\alpha_1}{\rightarrow} \dots
\stackrel{\alpha_{n-1}}{\rightarrow} \Sigma_n$ is a valid
derivation. We denote the set of all valid traces for the protocol
set $\Pi$ by $\tracesof{\Pi}$. When we consider the trace set
$\tracesof{\{P\}\cup\Pi}$ we say that $P$ runs in the context of
$\Pi$.

We reconstruct state information from a trace as follows. If
$\alpha_i$ is a run event from trace $\alpha$, then $\IK^{\alpha}_i$
is the intruder knowledge component $\IK$ of the state right
before the execution of $\alpha_i$.
Thus for all protocols $P$ and traces $\alpha \in \tracesof{P}$,
$\IK_0^{\alpha} = \IK_0$.

Next, we define a useful short hand.
Let $\Pi$ be a protocol set with $P \in \Pi$, and let $\alpha
\in \tracesof{\Pi}$. A \emph{cast} for $P$ in $\alpha$ is a map
$\cast:\domain(P) \rightarrow \runidset$ such that for some fixed
$\rho$, %
for every role $r \in \domain(P)$ there is a run event $\alpha_i =
((\cast(r),\rho,\anything), \createevent_\thelabel(r))$ with
$\thelabel \in P$. Intuitively, for a trace $\alpha$, a cast is an
assignment of runs to roles, which expresses the possibility that
these runs together form a session of the protocol. 
We denote the set of casts for $P$ and $\alpha$ by
$\castset{P}{\alpha}$.

\begin{exmp}
\label{ex:atrace} We illustrate the concept of a trace by
providing, in Figure~\ref{fig:nsltrace}, a possible trace of the \nsl'{} protocol from
Example~\ref{ex:nslprime}. This trace consists of the execution of
three runs. The first run is an instantiation of role $\irole$,
with instantiation $(1, \{ \irole \mapsto a, \rrole \mapsto b \},
  \{ \nonr \mapsto u, \nonrp \mapsto \undef \}
)$, where $a$ and $b$ are agents, $u$ is a nonce run term, and
$\undef$ means that no value has been assigned yet. The second run
instantiates role $\irole$ as $(2, \{ \irole \mapsto a,
\rrole \mapsto b \},
  \{ \nonr \mapsto v, \nonrp \mapsto \undef \}
)$, for nonce run term $v \neq u$. The third run is an
instantiation of the responder role $\rrole$, through $(3, \{
\irole \mapsto a, \rrole \mapsto b \},
  \{ \nonr \mapsto u, \nonip \mapsto \undef \}
)$.
Since runs 1 and 3 are instantiated such that they correspond,
they can be executed up to completion. In contrast, run 2 is
blocked. %

\begin{figure}[htb]
$$
\begin{tabular}{|l|l|}
\hline
instantiation & event \\
\hline
$(1,\rho, \{ \nonr \mapsto u, \nonrp \mapsto \undef \})$ &
  $\createevent_{1} (\irole) \seq$ \\
$(2,\rho, \{ \nonr \mapsto v, \nonrp \mapsto \undef \})$ &
  $\createevent_{1} (\irole) \seq$ \\
$(2,\rho, \{ \nonr \mapsto v, \nonrp \mapsto \undef \})$ &
  $\sendevent_{2} (\irole, \rrole, \enc{\nonip,\irole,\nonr}{\pkr}) \seq$ \\
$(1,\rho, \{ \nonr \mapsto u, \nonrp \mapsto \undef \})$ &
  $\sendevent_{2} (\irole, \rrole, \enc{\nonip,\irole,\nonr}{\pkr}) \seq$ \\
$(3,\rho, \{ \nonr \mapsto u, \nonip \mapsto \undef \})$ &
  $\createevent_{6} (\rrole) \seq$ \\
$(3,\rho, \{ \nonr \mapsto u, \nonip \mapsto \nonip\markid{1}\})$ &
  $\readevent_{2} (\irole, \rrole, \enc{\nonip,\irole,\nonr}{\pkr}) \seq$ \\
$(3,\rho, \{ \nonr \mapsto u, \nonip \mapsto \nonip\markid{1}\})$ &
  $\sendevent_{3} (\rrole, \irole, \enc{\nonip,\nonrp,\rrole}{\pki}) \seq$ \\
$(1,\rho, \{ \nonr \mapsto u, \nonrp \mapsto \nonrp\markid{3} \})$ &
  $\readevent_{3} (\rrole, \irole, \enc{\nonip,\nonrp,\rrole}{\pki}) \seq$ \\
$(1,\rho, \{ \nonr \mapsto u, \nonrp \mapsto \nonrp\markid{3} \})$ &
  $\sendevent_{4} (\irole, \rrole, \enc{\nonrp}{\pkr}) \seq$ \\
$(3,\rho, \{ \nonr \mapsto u, \nonip \mapsto \nonip\markid{1}\})$ &
  $\readevent_{4} (\irole, \rrole, \enc{\nonrp}{\pkr}) \seq$ \\
$(3,\rho, \{ \nonr \mapsto u, \nonip \mapsto \nonip\markid{1}\})$ &
  $\stopevent_{7} (\rrole) \seq$ \\
$(1,\rho, \{ \nonr \mapsto u, \nonrp \mapsto \nonrp\markid{3} \})$ &
  $\stopevent_{5} (\irole)$ \\
\hline
\end{tabular}
$$
\caption{Example trace of the \nsl'{} protocol
($\rho = \{ \irole \mapsto a, \rrole \mapsto b \})$.
} \label{fig:nsltrace}
\end{figure}

After execution of this trace the intruder knowledge $\IK$ is
extended with the information contained in the four send events
from the trace. Thus we have that $\IK$ is equal to:
\[
\closure{ \IK_0 \cup
  \{
    \enc{\nonip\markid{2},\irole,v}{\pkr},
    \enc{\nonip\markid{1},\irole,u}{\pkr},
    \enc{\nonip\markid{1},\nonrp\markid{3},\rrole}{\pki},
    \enc{\nonrp\markid{3}}{\pkr}
  \}
}
\]
There are two casts for the \nsl'{} protocol in this trace: $\{ i
\mapsto 1, r \mapsto 3 \}$ and $\{ i \mapsto 2, r \mapsto 3 \}$.
\end{exmp}

\subsection{Trace restrictions}
\label{sec-trace-restriction}
\label{after-the-model}

In Sections \ref{secprot} and \ref{runs} we have described the
semantics proposed by Cremers and Mauw in~\cite{CrMa04b}. In this and
the next section we define mechanisms for properly handling
parameters, and we add several new security properties to the
semantics, both important for protocol composition.

Note that there are essentially no restrictions on which values
parameters take in the semantics. We interpret parameters as input
to the protocols, and as such we need to specify where the
protocol gets its input from. We do this on the level of protocol sets
by specifying which protocols produce output and which
protocols are allowed to use this output as input. This
specification is done by means of trace restrictions.

When we study protocols in isolation, we do not want to consider how
the input is created, we only want to consider what properties
hold for the input. We model these properties using trace
restrictions.
While some trace restrictions are related or similar to
security properties, trace restrictions do not model protocol
security properties, but rather usage of the protocols.

A trace restriction is essentially a predicate on a trace set. We
use this predicate as a filter, selecting a subset of the trace
set.
\begin{defn}
  Let $\Pi$ be a protocol set, and let $\chi$ be a predicate on
  $\tracesof{\Pi}$. Then
  \begin{equation*}
    \restrictedtraces{\Pi}{\chi} = \{ \alpha \in \tracesof{\Pi} \mid
    \chi(\alpha) \} \text.
  \end{equation*}
\end{defn}
\iffalse
We also extend the notion of \emph{satisfies} in the obvious way
to these new trace sets by writing
$\restrictedsatisfies{\Pi}{\thelabel}{\chi}$.
\fi

In practice, a user first executing a
protocol $P$ and then executing a second protocol $Q$ might pass
values obtained from $P$ as input to the execution of $Q$.
In a trace, %
the mechanism for passing the
input value is encoded by initializing the
parameter for the $Q$-run of an agent in some role,
to a value produced by the $P$-run of the same agent acting
in the same role.

There are two variants to this use. Either the user inputs a value
generated by $P$ into one or more executions of the same role of $Q$,
or he inputs a value into just one execution of the $Q$ role. We
define two trace restrictions corresponding to these uses, labeled
$IO$ and $IO!$.

We say that a protocol $P$ \emph{establishes} an identifier $c$ if for
any trace $\alpha$ and every occurrence of events
$(inst,\stopevent_\thelabel(r))$, $\thelabel\in P$, in the trace, $inst$ is defined at $c$. This condition can be verified by a
syntactical analysis.

For the remainder of this section, let $\Pi$ be a protocol set,
let $\Pi' \subseteq \Pi$ be a set where every protocol establishes
$c$, and let $\Pi'' \subseteq \Pi$ be a protocol set where every
protocol has $d$ as a parameter in every role. Let $\alpha \in
\tracesof{\Pi}$, and let $I$ contain all subscripts of events in
$\alpha$ which correspond to a create event from $\Pi''$:
\begin{equation*}
  I = \{ i \mid \alpha_i = (inst_i, \createevent_\thelabel(r_i))
  \wedge \thelabel\in \Pi'' \} \text.
\end{equation*}
For each $i\in I$, define the sets
\begin{align*}
  A_i & = \{ j \mid \alpha_j = (inst_j, \createevent_\thelabel(r_j))
  \wedge \thelabel \in \Pi'' \wedge inst_j(d) = inst_i(d) \} \text{
    and} \\
  B_i & = \{ j \mid \alpha_j = (inst_j,end_\thelabel(r_i)) \wedge
  \thelabel \in \Pi' \wedge inst_j(c) = inst_i(d) \} \text.
\end{align*}
The set $A_i$ contains the runs that received the same input as
the run that was started in $\alpha_i$, and $B_i$ contains the set
of runs that could have produced this input. We also define the
following subsets:
\begin{align*}
  A_i' & = \{ j \in A_i \mid \forall r \in \roleset : inst_j(r) =
  inst_i(r) \} \text{ and} \\
  B_i' & = \{ j \in B_i \mid \forall r \in \roleset : inst_j(r) =
  inst_i(r) \} \text,
\end{align*}
where the equality is in the sense that either both are defined
and equal, or both are undefined. The set $A_i'$ is therefore the
subset of $A_i$ of which the runs have the same $\rho$ as
$\alpha_i$, that is, they believe they are communicating with the
same partners. The set $B_i'$ has a similar interpretation.

\begin{defn}
  Let $\Pi'\not=\emptyset, \Pi''$ be such that all protocols in $\Pi'
  \cup \Pi''$ have the same role set. We define the predicates
  \begin{equation*}
    \chi_{IO}(\alpha; \Pi', \Pi'', c, d) \Leftrightarrow \forall i \in I \,
    \exists j \in B_i' \, \forall k \in A_i' : j < k
  \end{equation*}
  and
  \begin{multline*}
    \chi_{IO!}(\alpha; \Pi', \Pi'', c, d) \Leftrightarrow \\ \forall i \in I
    \exists f: A_i' \rightarrow B_i' : \text{$f$ injective} \wedge (\forall
    j \in A_i' : f(j) < j) \text.
  \end{multline*}
\end{defn}

The trace restriction $\chi_{IO}(\alpha; \Pi', \Pi'', c, d)$ says that
a protocol set $\Pi''$ takes its input from a protocol set
$\Pi'$. As explained before, this means that any run of a role of
a protocol in $\Pi''$ initializes its input parameter $d$ to a
value that has been recorded in $c$ earlier in the trace, in a run
of the corresponding role of a protocol in $\Pi'$. (The
initialization is specified by defining $\sigma$ only at the input
parameters when the run is created.) In the stricter trace
restriction, $\chi_{IO!}(\alpha; \Pi', \Pi'', c, d)$ it is
required that at most one run of a role of a protocol in $\Pi''$
can take as its input a value produced in a run of the
corresponding role of a protocol in $\Pi'$. In order to ease
notation, when $\Pi = \Pi_1 \cup \Pi_2$, $\Pi'$ is a subset of
$\Pi_1$ and $Q \in \Pi_2$, instead of writing
$\restrictedtraces{\Pi}{\chi_{IO}(\cdot ; \Pi', Q, c, d)}$ and
$\restrictedtraces{\Pi}{\chi_{IO!}(\cdot ; \Pi', Q, c, d)}$, we
simply write $\tracesof{\ioout{\Pi_1}{c} \cup \ioin{\Pi_2}{c}{d}}$
and $\tracesof{\ioout{\Pi_1}{c} \cup \ioinunique{\Pi_2}{c}{d}}$
respectively.

Our main goal is to study security properties of component
protocols in isolation. When one protocol takes input from another
protocol, we do not want to include the second protocol in the
analysis. Our strategy is instead to specify preconditions on the
input to our protocol, sufficient for the protocol to achieve its
goals. We express these preconditions in terms of trace
restrictions.

Next, we define what it means for input to be secret. We emphasize
that the trace restriction makes no claim about what eventually
happens with the input. It may very well not remain secret.  But the
construction is such that it is \emph{possible} to keep the input
secret. As an example, consider the empty protocol with a claim for
secrecy of the input identifier. Under a secret trace restriction on
the input, the secrecy claim should always be satisfied. To achieve
this, we resort to a technical trick. The idea is that since $d$ is a
parameter in the run with run identifier $rid$, the run term
$d\markid{rid}$ is guaranteed by the semantics not to be known by the
adversary or any other execution at the start of the run. Note that
$d\markid{rid}$ should not be thought of as a locally generated nonce,
it is merely a convenient name for something generated elsewhere.

\begin{defn}
  We define the predicate
  \begin{equation*}
    \chi_{\secretclaim}(\alpha; \Pi'', d) \Leftrightarrow \forall i \in
    I :  j = \min A_i \wedge inst_j = (rid,\anything,\anything) \Rightarrow
    inst_i(d) = d\markid{rid} \text.
  \end{equation*}
\end{defn}

A somewhat stronger notion of secrecy is that of session key, where
we not only have secrecy, but also a notion of a session.

\begin{defn}
  Let the event $\alpha_i$ have instantiation $inst_i$, label
  $\thelabel_i$ and role $r_i$, with $\thelabel_i \in P_i$. We define
  the predicate
  \begin{align*}
    \chi_{\sessionclaim}(\alpha; \Pi', d) & \Leftrightarrow  \\
    \forall i \in I\ & : (
    \exists j \in A_i'  : inst_j = (rid,\anything,\anything) \wedge
    inst_i(d) = d\markid{rid}) \wedge \mbox{} \\
    & \mathrel{\hphantom{:}} (\exists f: A_i' \rightarrow \roleset :
    \text{$f$ injective} \wedge \mbox{} \\
    & \qquad\qquad \forall j \in A_i': \thelabel_j \in P_i \wedge f(j) = r_j )
  \end{align*}
\end{defn}

Finally, a much simpler concept is that the protocol takes its
input from the adversary. In this case, the idea is that the
protocol does not really care about where its input comes from,
just that it gets its input. We do not believe this is interesting
on its own, but it is a useful tool in analysis. One such example
is protocols where the chaining nonce is public, for instance NSL
variants using signatures instead of public key encryption.

\begin{defn}
  We define the predicate
  \begin{multline*}
    \chi_{\adversaryinput}(\alpha; \Pi', d) \Leftrightarrow \\ \forall 1
    \leq i \leq \length{\alpha} : \alpha_i = (inst,
    \createevent_\thelabel(r)) \wedge \thelabel \in \Pi' \Rightarrow
    inst(d) \in M^\alpha_i \text.
  \end{multline*}
\end{defn}

To simplify the notation, we denote these trace sets simply as
$\tracesof{\ioinsecret{\Pi}{d}}$,
$\tracesof{\ioinsession{\Pi}{d}}$ and
$\tracesof{\ioinadversary{\Pi}{d}}$.

\iffalse
Furthermore, whenever we consider
a protocol set $\Pi_1$ we allow unspecified hidden trace restrictions
to be active for that protocol set, as long as
those restrictions only concern protocols in $\Pi_1$, and no other
protocols.

As an example, when we talk about the trace set
$\tracesof{\ioout{\Pi_1}{c} \cup \ioin{\Pi_2}{c}{d}}$, we allow
hidden trace restrictions on $\Pi_1$ saying that $P \in \Pi_1$
produces input for $R \in \Pi_1$ (and $R$ could be one of the
protocols that produce $c$ as input to some protocol in $\Pi_2$).
However, a trace restriction saying that $R'$ in $\Pi_2$ takes
input from $P$ in $\Pi_1$ is not considered hidden.

\note{SR}{The two paragraphs above are still hard to understand for the reader, since the reader has by  now been exposed to many definitions and simplifications and hasn't seen any typical usage of these concepts. Is it necessary to point out that we have hidden trace restrictions? Where do we use them?}
\note{KG}{We don't. I've taken it out. But it must go back in when we
  write a book about this framework.}
\note{SM}{We are writing a book, aren't we? Or is it a PhD thesis?}
\note{CC}{Just that I'm clear about this, whose thesis are we writing?}
\fi

\subsection{Security properties}
\label{security-properties}

We have already introduced claim events in the trace model. Claim
events are not real protocol events, but markers we put in a trace to
indicate that a certain statement about the trace is supposed to hold.
We can, for instance, extend the role definition of $\nsl'(\irole)$
from Example~\ref{ex:nslprime} with claim event $\claimevent_{8}
(\irole, \secretclaim, \nonip )$, which contains the claim
$\secretclaim$. If an agent reaches this claim event during his
execution of role $\irole$, it is intended that an intruder will never
learn the value of his nonce.

The definition of security properties proceeds in three steps. First,
we define the general signature of a security property, which we
consider a predicate on protocol traces. Next, we express that such a
property is satisfied if it holds for all traces of the protocol.
Finally, we define a number of security properties, such as secrecy
and authentication.

In general, a security property is a predicate on the traces of a
protocol. So, given protocol $\Pi$ and a claim $cl$,
$f_{cl}(\Pi,\claimevent_\thelabel(r,cl,m))$ assigns truth values to
pairs $(inst,\alpha)$, where $\alpha$ is a trace of $\Pi$ and $inst$
is an instantiation of the variables in the claim event.

\begin{defn}
  A \emph{security property} is a function $f_{cl}$, $cl \in
  \claimset$, that associates with every pair $(\Pi,ev)$ of a protocol
  set $\Pi$ and a role claim event $ev$ with claim $cl$, a predicate on
  pairs of instantiations and traces:
  \begin{multline*}
    f_{cl}: \powerset{\protocols} \times \{ \claimevent_\thelabel(r,cl,m) \mid \thelabel
    \in \labelset, r \in \roleset, m \in \roletermset \} \rightarrow \\
    \bigcup_{\Pi\in\powerset{\protocols}} \{ \Inst \times \tracesof{\Pi} \rightarrow \{ \z{true}, \z{false} \} \} \text,
  \end{multline*}
  where $\powerset{\protocols}$ is the powerset of the set of all protocols.
\end{defn}

Note that a context is needed to evaluate an instantiation, and the
label $\thelabel$ on the role claim event determines this context.

A security property $f_{cl}$ is satisfied in protocol set $\Pi$ if
it yields true for all protocol traces containing this claim $cl$.
This is expressed in the following definition. However, we
have included the additional restriction that only
claims concerning sessions between trusted agents are evaluated. One cannot
expect, for instance, that a shared secret is really secret if one
of the communication partners is corrupted. Notice that this does
not rule out the possibility that a secret in a trusted session is
broken due to an interleaving of a session with untrusted partners.
Of course there are security properties for which this restriction
is not appropriate, but since the properties used in this paper all
have this restriction in common, it is included in the following
definition of satisfaction.

\begin{defn}
  Let $f_{cl}$ be a security property, $\Pi$ a protocol
  set, and $\thelabel$ the label of a claim event with claim
  $cl$. We say that $\Pi$ \emph{satisfies} the claim $\thelabel$,
  denoted by $\satisfies{\Pi}{\thelabel}$, if
  \begin{multline*}
    \forall \alpha\in\tracesof{\Pi} \forall i : \alpha_i = (inst,
    \claimevent_{\thelabel}(r, cl, m)) \Rightarrow \\
    f_{cl}(\Pi,\claimevent_{\thelabel}(r, cl, m))(inst,\alpha) \vee (inst =
    (\anything,\rho,\anything) \wedge \image(\rho) \not\subseteq
    \trustedagentset ) \text.
  \end{multline*}
\end{defn}

We extend this in the obvious way with trace restrictions and to sets
of claim event labels writing
$\restrictedsatisfies{\Pi}{\{\thelabel_i\}}{\chi}$.

As an example, we can claim secrecy for a particular role term $m$ by
inserting a suitable claim event into the protocol specification.
That claim event will translate into the following statement about a trace $\alpha$: The adversary never learns the run term $inst(m)$ in the trace $\alpha$.

\begin{defn}
  Let $\alpha \in \tracesof{\Pi}$.
  The security property $f_{\secretclaim}$ associates the protocol
  set $\Pi$ and the claim event $ev =
  \claimevent_\thelabel(r,\secretclaim,m)$ with the statement
  \begin{equation*}
    f_{\secretclaim}(\Pi,ev)(inst, \alpha) \Leftrightarrow
    inst(m) \not\in
    \IK^{\alpha}_{\length{\alpha}+1} \text,
  \end{equation*}
  where the initial intruder knowledge is determined from $\Pi$.
\end{defn}

We also need to express that a given run is part of a \emph{session}
for its protocol. We achieve this by requiring that no two runs of the
same role of the protocol have the same value for the \emph{session
identifier} (the argument). We call this property \emph{session
uniqueness}.

\begin{defn}
  The security property $f_\sessionuniqueclaim$ associates the protocol set
  $\Pi$ and the claim event $ev =
  \claimevent_\thelabel(r,\sessionuniqueclaim,m)$ with the statement
  \begin{align*}
    f_\sessionuniqueclaim(\Pi,ev)(inst,\alpha) & \Leftrightarrow
    \forall 1 \leq i, j \leq \length{\alpha}  : \\
    & (\alpha_i = (inst_i,
    \claimevent_{\thelabel}(r,\anything,\anything)) \wedge \mbox{} \\
    & \hphantom{(} \alpha_j =
    (inst_j,\claimevent_{\thelabel}(r,\anything,\anything)) \wedge
    \mbox{} \\
    & \hphantom{(} inst_i(m) = inst_j(m) = inst(m)) \Rightarrow i = j
    \text.
  \end{align*}
\end{defn}
\iffalse
Note that we sometimes want to evaluate $f_{cl}(\Pi,ev)(inst,\alpha)$
even if no run event with role event $ev$ appears in the trace
$\alpha$. (See $f_\sessionclaim$ later on for an example.) Therefore,
we do not require that the labeled claim events in the trace match
$ev$, only that they have the same label. The other security
properties are defined in a similar way.
\fi

Many protocols establish \emph{session keys}, and we identify three
requirements that a session key needs to satisfy:
\begin{enumerate}
\item The session key must be secret.
\item There must be a session, that is, one run of each role of the
  protocol must know the session key.
\item The key must act as a session identifier, that is, it must be
  unique across all runs of the same role of the same protocol.
\end{enumerate}
The first requirement is taken care of by the secrecy property and
the third requirement by the session-unique property. The second
requirement is taken care of by data agreement for the session key,
which we now define. The idea is that, for every other role in the
protocol, there must exist a run that has the same value for the
term at some event causally preceding the claiming event.
These runs must also agree on which agent executes which role, thus possibly forming a session
of the protocol.

\begin{defn}
  The security property $f_\dataagreeclaim$ associates the protocol set
  $\Pi$ and the claim event $ev =
  \claimevent_\thelabel(r,\dataagreeclaim,m)$, with $\thelabel \in P$
  with the statement
  \begin{align*}
    f_\dataagreeclaim (\Pi,ev)(inst, \alpha) & \Leftrightarrow \exists \cast
    \in \castset{P}{\alpha} : \\ %
    & \mathrel{\hphantom{\Leftrightarrow}} \quad \forall r' \in
    \domain(P)\ \exists j : %
    \alpha_j = (inst_j,\anything)
    \wedge \mbox{} \\
    & \mathrel{\hphantom{\Leftrightarrow}} \qquad inst_j =
    (\cast(r'),\anything,\anything) \wedge inst_j(m) = inst(m) \text.
  \end{align*}
\end{defn}

Note that causal precedence is implicitly required in this
definition. Given any trace with a claim event, we can create a new
trace by removing any event not causally preceding the claim event.
Hence, for the definition to be satisfied, there must be agreeing
events for all roles causally preceding the claim event.

Now we can define the session key claim.

\begin{defn}
  The security property $f_\sessionclaim$ associates the protocol set $\Pi$
  and the claim event $ev = \claimevent_\thelabel(r,\sessionclaim,m)$
  with the statement
  \begin{align*}
    f_\sessionclaim&(\Pi,ev)(inst, \alpha) \Leftrightarrow \\
    & f_\secretclaim(\Pi,
    \claimevent_\thelabel(r,\secretclaim,m))(inst, \alpha) \wedge
    \mbox{}
    \\
    & %
    f_\sessionuniqueclaim(\Pi,
    \claimevent_\thelabel(r,\sessionuniqueclaim,m))(inst, \alpha) \wedge
    \mbox{} \\
    & %
    f_\dataagreeclaim(\Pi,
    \claimevent_\thelabel(r,\dataagreeclaim,m))(inst, \alpha) \text.
  \end{align*}
\end{defn}

We also define a weaker session key claim, where we drop the
requirement about the agreement with (and therefore existence of) communication partners.

\begin{defn}
  The security property $f_\wsessionclaim$ associates the protocol set $\Pi$
  and the claim event $ev =
  \claimevent_\thelabel(r,\wsessionclaim,m)$ with the statement
  \begin{align*}
    f_\wsessionclaim&(\Pi,ev)(inst, \alpha) \Leftrightarrow \\
    & f_\secretclaim(\Pi,
    \claimevent_\thelabel(r,\secretclaim,m))(inst,\alpha) \wedge \mbox{}
    \\
    & %
    f_\sessionuniqueclaim(\Pi,
    \claimevent_\thelabel(r,\sessionuniqueclaim,m))(inst,\alpha) \text.
  \end{align*}
\end{defn}

Finally, we deal with authentication. Our preferred notion is
\emph{synchronization}~\cite{CrMaVi06}, a strong form of
authentication. A non-injective synchronization claim holds if there
are executions of the other protocol roles whose $\readevent$ and
$\sendevent$ events match the claiming execution's events, up to the
claim event. An even stronger notion of authentication,
\emph{injective synchronization}, holds if there is exactly one set of
executions of the other protocol roles such that the $\readevent$ and
$\sendevent$ events match the claiming execution's events, up to the
claim event.

\begin{defn}
  The security property $f_\nisynchclaim$ associates the protocol set $\Pi$
  and the claim event $ev =
  \claimevent_\thelabel(r,\nisynchclaim)$, $\thelabel \in P$ for some
  $P \in \Pi$ with the statement
  \begin{align*}
    f_\nisynchclaim(\Pi,ev)(inst,\alpha) & \Leftrightarrow \exists \cast
    \in \castset{P}{\alpha}, 1 \leq i \leq \length{\alpha} :
    \alpha_i = (inst, ev) \wedge \mbox{} \\
    & \forall r' \in \domain(P) \forall \thelabel' \in P(r) :
    \thelabel' \in P(r') \wedge \readevent_{\thelabel'} \before
    ev \wedge \mbox{} \\
    & \Rightarrow \exists j < k < i : \\
    & \quad \alpha_j = (inst'', \sendevent_{\thelabel'}(m)) \wedge
    \alpha_k = (inst''',\readevent_{\thelabel'}(m')) \wedge \mbox{} \\
    & \quad inst''(m) = inst'''(m') \wedge \mbox{} \\
    & \quad ((inst'' = (\cast(r),\anything,\anything) \wedge inst''' =
    (cast(r'),\anything,\anything)) \vee \mbox{} \\
    & \quad \hphantom{(} (inst'' = (\cast(r'),\anything,\anything)
    \wedge inst''' = (cast(r),\anything,\anything))) \text.
  \end{align*}
\end{defn}

\begin{defn}
  The security property $f_\isynchclaim$ associates the protocol set $\Pi$
  and the claim event $ev =
  \claimevent_\thelabel(r,\isynchclaim)$, $\thelabel \in P$ for some
  $P \in \Pi$ with the statement
  \begin{align*}
    f_\isynchclaim(\Pi,ev)(inst,\alpha) & \Leftrightarrow \exists! \cast
    \in \castset{P}{\alpha}, 1 \leq i \leq \length{\alpha} :
    \alpha_i = (inst, ev) \wedge \mbox{} \\
    & \forall r' \in \domain(P) \forall \thelabel' \in P(r) :
    \thelabel' \in P(r') \wedge \readevent_{\thelabel'} \before
    ev \wedge \mbox{} \\
    & \Rightarrow \exists j < k < i : \\
    & \quad \alpha_j = (inst'', \sendevent_{\thelabel'}(m)) \wedge
    \alpha_k = (inst''',\readevent_{\thelabel'}(m')) \wedge \mbox{} \\
    & \quad inst''(m) = inst'''(m') \wedge \mbox{} \\
    & \quad ((inst'' = (\cast(r),\anything,\anything) \wedge inst''' =
    (cast(r'),\anything,\anything)) \vee \mbox{} \\
    & \quad \hphantom{(} (inst'' = (\cast(r'),\anything,\anything)
    \wedge inst''' = (cast(r),\anything,\anything))) \text.
  \end{align*}
\end{defn}

Certain security properties can be evaluated by merely looking at
the events in a trace that belong to the protocol in which the
claim was made. This class of properties are called
protocol-centric, and as we will see, we can prove theorems that
apply to all properties in this class. Since authentication properties
are concerned with the occurrence of events of the given protocol,
they are typical members of this class.

\begin{defn}
  Let $P$ be a protocol, and $\Pi$ a protocol set. Denote by $\pi_P$
  and $\pi_\Pi$ the maps on traces that remove any protocol event
  that does not belong to $P$ or a protocol in $\Pi$, respectively.
\end{defn}

Let $s$ be any bijection on the set of nonce run terms $\{
c\markid{rid} \mid c \in \idset, rid \in \runidset \} \cup \IT$.
This is basically a renaming of nonce run terms. Any such
bijection can be naturally extended to a bijection on the set of run terms.
For any trace $\alpha$, we define $s(\alpha)$ to be the trace
where every instantiation $inst$ is replaced with $s \circ inst$.

\begin{defn}
  We say that a security property $f_{cl}$ is
  \emph{protocol-centric} if for any $(\Pi,ev)$ such that $f_{cl}$ is
  defined and $ev$ belongs to a protocol $P \in \Pi$, and for any
  renaming $s$ on nonce run terms,
  \begin{multline*}
    \forall \alpha, \alpha' \forall inst \in Inst: s(\pi_P(\alpha)) =
    \pi_P(\alpha') \\ \Rightarrow f_{cl}(\Pi,ev)(inst,\alpha) =
    f_{cl}(\Pi,ev)(s\circ inst,\alpha') \text.
  \end{multline*}
\end{defn}

(The renaming $s$ is included in this definition for technical
reasons.)

We observe that $\sessionuniqueclaim$, $\dataagreeclaim$,
$\nisynchclaim$ and $\isynchclaim$ are protocol-centric, while
$\secretclaim$ and therefore $\sessionclaim$ are not.

\section{Framework for reasoning}
\label{sec:reasoning}

Automatically proving large protocols secure is computationally
challenging. If the protocol can be split into a sequential composition of
several subprotocols, one approach to verification is to
analyze the subprotocols in a common context.
Properties proved for
each of the subprotocols running in this multi-protocol context can
often be used to deduce properties for the composed protocol.
Note that some authors consider protocols running in a multi-protocol
context to be composed in parallel. We use ``composition''
only about operations that combine two or more protocol objects into a
new protocol object. The only form of composition considered in this
paper is sequential composition. %
Unfortunately, automatically verifying protocol properties in such a
multi-protocol context is not computationally easier than analysing the
composed protocol itself. The ideal is to study each subprotocol in
perfect isolation, without consideration of any other protocols. Our
approach is to study under which conditions protocols running in
parallel can be shown not to interfere with each other, such that
results obtained by analysis in isolation will be valid for a
multi-protocol context.

Our approach is to use a very strong, but efficiently verifiable
notion of independence between
protocols. We show how to design protocols to ensure such independence
without incurring any significant performance penalty. %
We then prove a number of theorems showing
conditions under which protocols can run in a multi-protocol context without interfering with each other.
Finally, we define sequential protocol composition (similar to the one
in PCL) and show how
properties of subprotocols can be combined to give security
properties for the composed protocols.

We use the formal model described in Section~\ref{sec:framework}, with
two restrictions:
\begin{enumerate}
\item We require that the partial function $\sigma$ in instantiations
  assigns only nonce run terms to variables and parameters.

\item We restrict ourselves to protocols that use secret long-term
  keys only as keys, never as content of messages.
\end{enumerate}

The first restriction is essential for the
notion of strong independence, defined in the next section. We refer to
Section~\ref{sec:conclusion} for a discussion of the implications. The
second restriction could be lifted, but this
would subtly complicate analysis, since the use of long-term secret keys
becomes much harder to predict.

\subsection{Independence}

We say that two protocols are independent if no encryption term
produced by the first protocol running in the context of the
second protocol will be decrypted or verified by the second
protocol, and vice versa. Formally:

\begin{defn}
  Let $\Pi_1,\Pi_2$ be two disjoint protocol sets, and let $\chi$ be a
  (possibly empty) trace restriction. We say that \emph{$\Pi_1$ and
    $\Pi_2$ are independent in the context of $\chi$}, denoted
  $\independentc{\Pi_1}{\Pi_2}{\chi}$ (alternatively if $\chi$ is
  empty, \emph{$\Pi_1$ and $\Pi_2$ are independent}, denoted
  $\independent{\Pi_1}{\Pi_2}$), if
  \begin{align*}
    \forall \alpha & \in \restrictedtraces{\Pi_1 \cup \Pi_2}{\chi}
    \,\, \forall x,y,x',y' \in \roletermset: \\
    \Bigl( & \alpha_i = (inst,\sendevent_\thelabel(m)) \wedge \bigl(\alpha_j =
    (inst',\readevent_{\thelabel'}(m')) \vee \alpha_j =
    (inst',\sendevent_{\thelabel'}(m')) \vee \mbox{} \\
    & \hphantom{\alpha_i = (inst,\sendevent_\thelabel(m)) \wedge (} \alpha_j =
    (inst',\claimevent_{\thelabel'}(\anything,\anything,m'))\bigr) \wedge \mbox{} \\
    & \quad \bigl( \enc{x}{y} \subterm m \wedge
    \enc{x'}{y'} \subterm m' \wedge inst(\enc{x}{y}) =
    inst'(\enc{x'}{y'}) \bigr) \Bigr) \\
    & \Rightarrow (\thelabel,\thelabel' \in \Pi_1 \vee
    \thelabel,\thelabel' \in \Pi_2) \text.
  \end{align*}
\end{defn}

\iffalse
\begin{thm}
  \label{thm:io.independent.implies.no-io.independent}
  Suppose $\ioout{\Pi_1}{c}$ and $\ioin{\Pi_2}{c}{d}$ are two
  independent protocol sets. Then $\Pi_1$ and $\ioinsession{\Pi_2}{d}$
  are independent.
\end{thm}
\begin{proof}
  Let $\alpha \in \tracesof{\Pi_1 \cup \ioinsession{\Pi_2}{d}}$.
  Note that we can always generate complete runs of a protocol where
  the adversary does not interfere. Before the first create event of
  every session initializing the parameter to the value $t$, we insert

  \note{KG}{Idea: If $\Pi_2$ cannot, when it gets some possibly useful
    information from $\Pi_2$, send or read a term that $\Pi_1$ will
    send or accept, then it cannot do it when it doesn't get that
    information. This should be obvious! But how to formulate a
    proof?}

\note{SR}{How about mapping $Tr(\Pi_1,\ioinsecret{\Pi_2}{d})$
into $Tr(\ioout{\Pi_1}{c},\ioin{\Pi_2}{c}{d})$ by inserting clean $\Pi_1$ runs before every $\Pi_2$ run. If they are not independent in the domain of the map they will not be independent in the image.
}
\note{KG}{It would work for the session trace restriction, but secret
  allows more roles to get the input than session does. A single clean
  run may not be enough to feed everyone with a value.}
\end{proof}
\fi

In general, proving independence is a non-trivial problem, but for many protocol sets it is easy in the sense that the protocol sets satisfy an even stronger notion of independence.
 We say that two protocol sets are strongly independent if they have no
encryptions of the same form.
Unlike independence, strong independence can be easily verified at
the syntactical level, and it implies independence. Note that
different protocols can use the same cryptographic keys and still be
strongly independent, and thus independent.

\begin{defn}
  Let $\Pi_0$ and $\Pi_1$ be two disjoint protocol sets. We say that
  \emph{$\Pi_0$ and $\Pi_1$ are strongly independent}, denoted
  $\sindependent{\Pi_0}{\Pi_1}$, if for any $b\in \{0,1\}$, any role
  specification $(elist\cdot \sendevent(m)\cdot elist',\idtyper)$ in a
  protocol in $\Pi_b$, any role terms $x,y$, any
  role specifications $(elist''\cdot \sendevent(m') \cdot
  \elist''',\idtyper')$, $(elist''\cdot \readevent(m') \cdot
  \elist''',\idtyper')$ or $(elist''\cdot \claimevent(r,c,m') \cdot
  \elist''',\idtyper')$ in protocols of $\Pi_{1-b}$, any map $s$ on
  the set $\idset$ and any map $s'$ on the set $\roleset$,
  \begin{equation*}
    \enc{x}{y} \subterm m \Rightarrow \enc{s(s'(x))}{s(s'(y))}
    \not\subterm m' \text.
  \end{equation*}
\end{defn}

Note that any map $s$ on identifiers and $s'$ on roles naturally induce
maps on the set of role terms and we identify these maps with $s$ and
$s'$.
Also note that since strong independence is a syntactical
property, there is no need to consider trace restrictions.

\begin{thm}
  \label{thm:strong.independence}
  If two protocol sets $\Pi_1$ and $\Pi_2$ are strongly independent,
  then they are independent.
\end{thm}
\begin{proof}
  Obvious from the fact that only nonce run terms are assigned to
  variables and parameters.
\end{proof}

The notion of strongly independent protocols is obviously very
strong, and there are many independent protocols that are not
strongly independent. However, it is possible to verify strong
independence by a simple syntactical check on the protocols.
Obviously, strong independence may not be useful for analysing some
existing protocols, but it does cover many deployed protocols (see
Section~\ref{sec:wimax}).

One way to achieve strong independence is to use separate key
infrastructures for every protocol. Unfortunately, this is
expensive and wasteful. A more practical way to get strong
independence is through \emph{protocol tags}. Every protocol is
given a unique tag which is embedded into every ciphertext the
protocol makes. This trivially implies strong independence.

Protocol tags may be a desirable approach for protocol design, since
they typically require no extra bandwidth, and only modest extra
computational effort.

Modern signature schemes typically process the message to be
signed with a hash function, create a signature tag, and attach
the tag to the message. Adding a protocol tag of reasonable length
(say 128 bits) to the message will usually result in a minor
increase in the cost of computing the hash function. Since the
signature is simply attached to the message, and both signer and
verifier know the protocol tag, there is no need to actually
transmit the protocol tag. The signer can remove the tag from the
message before transmitting, the verifier puts the tag back in
before verifying. A signature made by one protocol will not pass
the verification by a second protocol.

Modern encryption schemes typically allow for part of the message
to be left unencrypted but authenticated. Again, if we include the
protocol tag in the unencrypted part, the encrypter can remove the
tag before transmission and the decrypter can insert the tag prior
to decryption. Typically, a protocol tag of reasonable length will
result in a minor increase in the cost of authentication.

As for hash-like function evaluations, most cryptographically
interesting functions either allow the protocol tag to be inserted
into the function evaluation, or allow cryptographic separation by
choosing distinct parameters. The computational cost of most such
measures are expected to be modest.

To summarize, protocol tags typically have no bandwidth cost and
modest computational cost. This suggests that protocol tagging is
a viable and sensible strategy for protocol design.

\subsection{Multi-protocol environments}

Once we have independence, we are ready to prove that any protocol
remains correct in the presence of independent protocols. The
general idea for proving all of the results in this section is to
define maps between trace sets, and then argue that the predicate
derived from a claim statement remains unchanged under this map.

The next theorem says that if one protocol set keeps something
secret, it will keep it secret even in the presence of a second,
but independent, protocol set.

\begin{thm}
\label{thm:basic.secret}
  Let $\Pi_1$ and $\Pi_2$ be two independent protocol sets. Let
  $\thelabel$ be the label of some secret claim event in $\Pi_1$. Then
  \begin{equation*}
    \satisfies{\Pi_1}{\thelabel} \Rightarrow \satisfies{\Pi_1\cup
      \Pi_2}{\thelabel} \text.
  \end{equation*}
\end{thm}
\begin{proof}
  Let $S$ be the set of nonce run terms in $\alpha$ originating in
  runs of roles from protocols in $\Pi_2$. Let $s:S \rightarrow \IT$
  be an injection such that no term in the image of $s$ appears in
  $\alpha$. We can extend $s$ to the set of nonce run terms by letting
  $s$ be the identity where it is not already defined. This map can
  then be extended naturally to a renaming map on the set of run
  terms.

  We construct a new trace $\alpha'$ from $\alpha$
  by removing any events belonging to runs of roles from protocols in
  $\Pi_2$, and replacing any other event $(inst,ev)$ by $(s\circ
  inst,ev)$. (Since $s$ renames only nonce run terms, the
  composition $s\circ inst$ may affect only the $\sigma$ function
  of $inst$.)
  Note that by independence, if any nonce run term originates in a run of
  a role of a protocol in $\Pi_2$, the only way a run of a role of a
  protocol in $\Pi_1$ will read that nonce run term is if the adversary
  also knows that nonce run term.  From the semantics, we have that a nonce of
  $\Pi_2$ can only occur in a run of a role of $\Pi_1$ as a subterm of an
  instantiated variable. Therefore, if we replace the nonce run term by an
  attacker-generated nonce run term (such that the type constraints on the containing
  variable are met), the trace will still be valid even after the $\Pi_2$-events
  are removed. This means that $\alpha' \in \tracesof{\Pi_1}$.

  There is always a canonical choice of injection $s$ (given the
  well-ordering on the nonce run terms induced by the natural
  numbers), and this gives us a map
  \begin{equation}
    \label{eq:projection.with.some.replacement}
    \tau : \tracesof{\Pi_1 \cup \Pi_2} \rightarrow \tracesof{\Pi_1} \text.
  \end{equation}
  Note that $\tau = s \circ \pi_{\Pi_1}$.

  Now consider a run term $t$ claimed secret in $\alpha$. In
  $\alpha'$, the corresponding run term is $s(t)$, and we know that
  this is secret. We first determine why $s(t)$ is secret, and we may
  as well assume that $s(t)$ is a non-tuple run term. If $s(t)$ has
  the form $f(u)$ for some function $f$ and run term $u$, then $t$ is
  secret by assumption. If $s(t)$ is a nonce run term, we know that first of
  all $s(t)$ must originate in a run of a role of a protocol in
  $\Pi_1$. Second, every time it appears in a sent run term, it must
  be inside an encryption term. By independence, no $\Pi_2$-run will
  decrypt that ciphertext. Therefore, $t$ must be secret in $\alpha$.
  Otherwise, $t$ must have the form $\enc{u}{v}$ for some run terms $u$
  and $v$. If $s(u)$ is secret, we must show that $u$ is secret. We
  consider $u$ instead of $t$ and return to the start of the argument.
  Otherwise, $s(v)$ must be secret. By independence, it is sufficient
  to show that $v$ is secret, so we consider $v$ instead of $t$ and
  return to the start of the argument.

  Since terms cannot be infinitely nested, this argument chain must
  eventually stop, and in the process prove that $t$ is secret in
  $\alpha$. This concludes the proof.

  \iffalse We use the map $\tau$ from
  \eqref{eq:projection.with.some.replacement}.  Let $\alpha \in
  \tracesof{\Pi_1 \cup \Pi_2}$ and let $s$ be such that $\tau(\alpha)
  = (s \circ \pi_{\Pi_1})(\alpha)$.

  Consider a run term $t$ claimed secret in $\alpha$. In
  $\alpha'$, the corresponding run term is $s(t)$, and we know that
  this is secret. We first determine why $s(t)$ is secret, and we may
  as well assume that $s(t)$ is a non-tuple run term. If $s(t)$ has
  the form $f(u)$ for some function $f$ and run term $u$, then $t$ is
  secret by assumption. If $s(t)$ is a nonce run term, we know that first of
  all $s(t)$ must originate in a run of a role of a protocol in
  $\Pi_1$. Second, every time it appears in a sent run term, it must
  be inside an encryption term. By independence, no $\Pi_2$-run will
  decrypt that ciphertext. Therefore, $t$ must be secret in $\alpha$.
  Otherwise, $t$ must have the form $\enc{u}{v}$ for some run terms $u$
  and $v$. If $s(u)$ is secret, we must show that $u$ is secret. We
  consider $u$ instead of $t$ and return to the start of the argument.
  Otherwise, $s(v)$ must be secret. By independence, it is sufficient
  to show that $v$ is secret, so we consider $v$ instead of $t$ and
  return to the start of the argument.

  Since terms cannot be infinitely nested, this argument chain must
  eventually stop, and in the process prove that $t$ is secret in
  $\alpha$. This concludes the proof.\fi
\end{proof}

If secrecy of some nonce is not important for satisfying some secrecy
claim in some protocol set, then passing the nonce to an independent
protocol set will not compromise the secrecy claim. The intuition is
that the worst an independent protocol can do is to reveal the nonce
to the intruder, and therefore we only need to analyse what happens in
that case.

\begin{defn}
  Let $P$ be a protocol establishing $c$. Then $\iooutreveal{P}{c}$ is
  the protocol
  \begin{equation*}
    \iooutreveal{P}{c} = \{ r \mapsto s \cdot
    \sendevent(r,\nu_{r_0,r_1}(r),c) \cdot \stopevent(r) \mid P(r) = s
    \cdot \stopevent(r) \} \text,
  \end{equation*}
  where $r_0$ and $r_1$ are two distinct roles of $P$ and
  $\nu_{r_0,r_1}(r)$ is $r_0$ when $r\not=r_0$, otherwise $r_1$.
\end{defn}

We extend this notation to protocol sets in the obvious way, writing
$\iooutreveal{\Pi}{c}$.

\begin{thm}
 \label{thm:secret.non-secret.parameter.passing}
  Let $\iooutreveal{\Pi_1}{c}$ and $\ioinadversary{\Pi_2}{d}$ be two
  independent protocol sets and let $\thelabel$ be the label of some
  $\secretclaim$ claim event in $\Pi_1$. Then
  \begin{equation*}
    \satisfies{\iooutreveal{\Pi_1}{c}}{\thelabel} \Rightarrow
    \satisfies{\ioout{\Pi_1}{c} \cup \ioin{\Pi_2}{c}{d}}{\thelabel} \text.
  \end{equation*}
\end{thm}
\begin{proof}
  It is clear that $\tracesof{\ioout{\Pi_1}{c} \cup
    \ioin{\Pi_2}{c}{d}}$ embeds naturally in
  $\tracesof{\iooutreveal{\Pi_1}{c} \cup \ioinadversary{\Pi_2}{d}}$.
  Furthermore, under this embedding the intruder knowledge is strictly
  increased. By Theorem~\ref{thm:basic.secret}, the secrecy claim
  holds in the latter trace set. It must therefore also hold in the
  former trace set and the theorem is proven.
\end{proof}

If secrecy of some nonce may be important for some secrecy claim,
then passing the nonce to an independent protocol set that
preserves the secrecy of its input will not compromise the secrecy
claim. (Note that the secrecy claims in the second protocol must
be positioned at the start of the role. Otherwise, the protocol
would be allowed to compromise the secrecy of the input as long as
none of its roles reaches its secrecy claim.)

\begin{thm}
  \label{thm:secret.secret.passing}
  Let $\ioout{\Pi_1}{c}$ and $\ioin{\Pi_2}{c}{d}$ be two protocol
  sets, let $\Pi_1'\subseteq\Pi_1$ be a set of protocols establishing
  $c$ and $\Pi_2' \subseteq \Pi_2$ be a set of protocols with $d$ as a
  parameter. Let $\thelabel \in \Pi_1 \cup \Pi_2$ be the label of some
  $\secretclaim$ claim event. If $\Pi_1$ and $\Pi_2$ are independent
  under the trace restrictions $\chi_{IO}(\anything ;
  \Pi_1',\Pi_2',c,d)$ and $\chi_\secretclaim(\anything ; \Pi_2', d)$,
  then
  \begin{equation*}
    \satisfies{\Pi_1}{\{\thelabel_i\} \cup \{\thelabel\}} \wedge
    \satisfies{\ioinsecret{\Pi_2}{d}}{\{\thelabel_i'\} \cup \{\thelabel\}} \Rightarrow
  \satisfies{\ioout{\Pi_1}{c} \cup
    \ioin{\Pi_2}{c}{d}}{\thelabel} \text.
  \end{equation*}
  Here we assume that $\{\thelabel_i\}$ are the labels of claim events
  $\{\claimevent_{\thelabel_i}(r_i,\secretclaim,c)\}$ in every role of
  every protocol in $\Pi_1$ that establishes $c$ and the
  $\{\thelabel_i'\}$ are labels of claim events
  $\{\claimevent_{\thelabel_i'}(r_i',\secretclaim,d)\}$ in every role
  of every protocol in $\Pi_2$ that has $d$ as a parameter. The event
  $\claimevent_{\thelabel_i'}(r_i',\secretclaim,d)$ is assumed to occur
  before any $\sendevent$ or $\readevent$ event in the role specification.
\end{thm}

\begin{proof}
  Let $\alpha\in\tracesof{\ioout{\Pi_1}{c} \cup
    \ioin{\Pi_2}{c}{d}}$ be a trace. Let $I$ be the set of indexes such that
  $\alpha_i = (inst_i,\createevent(r))$ for some role $r$ of a
  protocol $Q$ in $\Pi_2$ that takes $c$ as input for the parameter
  $d$. For each $i$, define the set
  \begin{equation*}
    A_i = \{ j \in I \mid inst_j(d) = inst_i(d) \} \text.
  \end{equation*}
  Let $rid_i$ be such that $\alpha_{\min A_i} =
  ((rid_i,\anything,\anything),\anything)$. Note that for any $i \in
  I$, the nonce run term $d\markid{rid_i}$ never appears in $\alpha$. Let
  $S = \{ inst_i(d) \mid i \in I \}$ and $S' = \{ d \markid{rid_i} \mid
  i \in I \}$, and let $s$ be the substitution that maps $inst_i(d)$
  to $d\markid{rid_i}$ for all $i$ in $I$.

  We construct a new trace $\alpha'$ from $\alpha$
  by replacing any event $(inst, ev)$ that belongs to $\Pi_2$ by the
  event $(s\circ inst, ev)$.

  We claim that $\alpha' \in \tracesof{\Pi_1 \cup
    \ioinsecret{\Pi_2}{d}}$, and we get a map
  \begin{equation}
    \label{eq:replacing.input.by.secret}
    \tau : \tracesof{\ioout{\Pi_1}{c} \cup \ioin{\Pi_2}{c}{d}}
    \rightarrow \tracesof{\Pi_1 \cup \ioinsecret{\Pi_2}{d}} \text,
  \end{equation}
  along with a natural bijection
  \begin{equation*}
    \theta: \IK^\alpha \rightarrow \IK^{\alpha'} \text.
  \end{equation*}

We will prove the claim and construct the map $\theta$ by induction.
The theorem will then follow from a simple observation.

  Suppose $\alpha_0'\cdots\alpha_{j-1}'$ is a valid trace. The subterm
  relation $\subterm$ defines a partial ordering on $\IK_j^{\alpha}$
  and $\IK_j^{\alpha'}$. Let $U_j$ and $U_j'$ be
  the minimal elements
  of these sets, together with the
  terms that cannot be inferred from smaller terms:
  \begin{align*}
    U_j &= \{ x \in \IK_j^{\alpha} \mid \text{$t$ $\subterm$-minimal}
    \} \cup \{ \enc{t}{t'} \in \IK_j^{\alpha} \mid t \not\in
    \IK_j^{\alpha} \vee t' \not\in \IK_j^{\alpha} \} \text, \\
    U_j' &= \{ x \in \IK_j^{\alpha'} \mid \text{$t$ $\subterm$-minimal}
    \} \cup \{ \enc{t}{t'} \in \IK_j^{\alpha'} \mid t \not\in
    \IK_j^{\alpha'} \vee t' \not\in \IK_j^{\alpha'} \} \text.
  \end{align*}
  Note that $\IK_j^{\alpha} = \closure{U_j}$ and $\IK_j^{\alpha'} =
  \closure{U_j'}$.
  Also note that any encryption term in $U_j$ and $U_j'$ must originate from
  some $\sendevent$ event.

  Define the sets
  \begin{equation*}
    V_j = \{ t \in U_j \mid \exists t' \in S : t' \subterm t \} \text{
      and } V_j' = \{ t \in U_j' \mid \exists t' \in S \cup s(S) : t'
    \subterm t \} \text.
  \end{equation*}
  Let $V_{j,1}$ and $V_{j,2}$ be the subsets of $V_j$ of elements
  originating in $\Pi_1$ and $\Pi_2$, respectively. Let
  \begin{align*}
    V_{j,1}' &= \{ t \in U_j' \mid \exists t' \in S : t' \subterm t \} \\
    V_{j,2}' &= \{ t \in U_j' \mid \exists t' \in s(S) : t' \subterm t \}
  \end{align*}
  Let $W_j = U_j \setminus V_j$ and $W_j' = U_j' \setminus V_j'$.

  We can now prove the claim and construct the map $\theta$ by
  induction on $j$. The induction hypothesis is that
  $\alpha_0'\cdots\alpha_{j-1}'$ is a valid trace and the following
  two properties hold
  for the structure of the intruder knowledge:
  \begin{enumerate}
  \item $S \cap V_j = \emptyset$ and $(S\cup s(S)) \cap V_j' =
    \emptyset$.
  \item There exists a bijection $\theta: U_j \rightarrow U_j'$ such
    that $\theta$ restricted to $W_j \cup V_{j,1}$ is the identity
    map, and $\theta$ restricted to $V_{j,2}$ corresponds to the map
    induced by the substitution $s$.
  \end{enumerate}
  Note that the bijection $\theta$ extends to a bijection $\theta:
  \IK_j^{\alpha} \rightarrow \IK_j^{\alpha'}$.

The induction basis is
  trivially satisfied for the empty trace and easy to verify for $\alpha_0'$.

  The trace restriction on $d$ is satisfied by design, so if
  $\alpha_j$ is a $\createevent$ event, $\alpha_0'\cdots\alpha_j'$ is
  a valid trace. Also, the same substitution is applied to all events
  in a run, so if $\alpha_j$ is a $\sendevent$, $\claimevent$ or
  $\stopevent$ event, $\alpha_0'\cdots\alpha_j'$ is a valid trace,
  because the instantiations will be consistent with the run
  (basically the instantiation of the last event).

  Next, we consider a $\readevent$ event $\alpha_j =
  (inst,\readevent_\thelabel(m))$, $\alpha_j' =
  (inst',\readevent_\thelabel(m))$. We must prove that $inst'(m) \in
  \IK_j^{\alpha'}$. We know that $inst(m) \in \IK_j^\alpha$. If
  $inst(m) \in \closure{W_j}$ we are done, and
  $\alpha_0'\cdots\alpha_j'$ is a valid trace.

  By independence we know that $V_{j,1} \cap V_{j,2} = \emptyset$.
  Again by independence, if $\thelabel \in \Pi_b$, then any run term
  in $V_j$ that is a subterm of $inst(m)$ is also in $V_{j,b}$, and we
  get that $inst(m)$ is in the closure of $V_{j,b} \cup W_j$. Note
  that $V_{j,1}' \cup W_j' = V_{j,1} \cup W_j$ and $V_{j,2}' \cup W_j'
  = s(V_{j,2} \cup W_j)$. If $\thelabel\in\Pi_1$ we have that
  $inst'(m) = inst(m) \in \IK_j^{\alpha'}$. If $\thelabel\in\Pi_2$ we
  must have that $inst'(m) = s(inst(m)) \in \IK_j^{\alpha'}$.
  Therefore, under the induction hypothesis,
  $\alpha_0'\cdots\alpha_j'$ is a valid trace.

  We finish the inductive step by showing that (1) and (2) are also
  satisfied after the $j$th event. We need only consider the
  event $\alpha_j = (inst,\sendevent_\thelabel(m))$, $\alpha_j' =
  (inst',\sendevent_\thelabel(m))$. The only interesting inference
  rule is $(\enc{t}{t'},t') \Rightarrow t$, and we will show that the
  structure is unchanged by decryptions, up to some trivial rewriting.

  For any set of run terms $T$, let $rcl(T)$ be the smallest set of
  run terms containing $T$ that is closed under tuple creation and
  dissolution, encryption with known keys and removing signatures.
  This is the restricted closure, closure without decryptions. Note
  that $\IK_j^{\alpha} = rcl(U_j)$.

  Under the induction hypothesis for $U$ and $U'$, if we augment
  $U$ by decrypting a run term $\enc{t}{t'}$, where $t' \in rcl(U)$, then
  we can augment $U'$ by decrypting the run term
  $\theta(\enc{t}{t'})$, since $\theta(t') \in rcl(U')$.  If
  $\theta(t) = t$, then clearly we augment $U$ and $U'$ in the same
  way.  Likewise, if $\theta(t) \not= t$,
  then $\theta(t)$ and $t$ are equal up to
  substitution by $s$, and $U$ and $U'$ are augmented in the same
  way, up to substitution. This means that $S \cap U = \emptyset$,
  because we know that $(S \cup s(S)) \cap U' = \emptyset$.  The maps
  can therefore be extended, and (1) and (2) still hold true after
  augmentation. Finally, if some of the elements in $U$ are no longer
  minimal and are not encryptions that should be preserved, then the
  corresponding elements in $U'$ will no longer be minimal, nor be
  encryptions that should be preserved. The other direction also
  holds. Therefore, we can discard all superfluous elements.

  The list $U_{j+1}$ can be reached from $U_j$ by adding the run terms
  obtained from the send event to the list, then performing a finite
  sequence of decryption operations, then possibly discarding some
  elements from the list. The above argument shows that the same
  operations (up to substitution) will turn $U_j'$ into $U_{j+1}'$ in
  such a way that (1) and (2) still hold for $U_{j+1}$ and
  $U_{j+1}'$. This completes the inductive step.

  To complete the proof of the theorem, we first observe that for any secrecy claim
  event $\alpha_j =
  (inst,\claimevent_\thelabel(\anything,\anything,m))$ there is $\alpha_j'
  = (inst',\claimevent_\thelabel(\anything,\anything,m))$ and
note that $\satisfies{\Pi_1 \cup
    \ioinsecret{\Pi_2}{d}}{\thelabel}$ is true by
  Theorem~\ref{thm:basic.secret},
thus $inst'(m) \not\in \IK^{\alpha'}$. Next, if $inst(m) \in
  \IK^{\alpha}$, then $\theta(inst(m))$ would be defined and equal to
  $inst'(m)$, contradicting $inst'(m) \not\in \IK^{\alpha'}$.
We conclude that $inst(m) \not\in \IK^{\alpha}$ and the
  secrecy claim holds.
\end{proof}

The following theorem states that protocol-centric claims
remain valid if we execute a protocol in the context of another
protocol that is independent of the first.

\begin{thm}
  \label{thm:basic.protocol-centric}
  Let $\Pi_1$ and $\Pi_2$ be two independent protocol sets, and let
  $\thelabel$ be the label of some claim event in $\Pi_1$. If the
  security property associated with $\thelabel$ is protocol-centric,
  then
  \begin{equation*}
    \satisfies{\Pi_1}{\thelabel}
    \Rightarrow \satisfies{\Pi_1 \cup \Pi_2}{\thelabel} \text.
  \end{equation*}
\end{thm}
\begin{proof}
  Since $\Pi_1$ and $\Pi_2$ are independent, we can use the trace map
  $\tau$ from \eqref{eq:projection.with.some.replacement} on page~\pageref{eq:projection.with.some.replacement}. Let $P$ be
  the protocol where the claim event with label $\thelabel$ appears.
  By construction of the $\tau$ map, we have  $s(\pi_P(\alpha))
  = \pi_P(\tau(\alpha))$,
  for some substitution $s$. Since the claim is protocol-centric, we
  are done.
\end{proof}

\iffalse
Protocol-centric claims even remain valid if there is (non-secret)
transfer of information from the first to the second protocol.

\begin{thm}
  \label{thm:protocol-centric.non-secret.passing}
  Let $\iooutreveal{\Pi_1}{c}$ and $\ioinadversary{\Pi_2}{d}$ be two
  independent protocol sets, and let $\thelabel$ be the label of some
  claim event in $\Pi_1$. If the security property associated with
  $\thelabel$ is protocol-centric, then
  \begin{equation*}
    \satisfies{\iooutreveal{\Pi_1}{c}}{\thelabel} \Rightarrow
    \satisfies{\ioout{\Pi_1}{c} \cup \ioin{\Pi_2}{c}{d}}{\thelabel} \text.
  \end{equation*}
\end{thm}
\begin{proof}
  If $\alpha \in \tracesof{\ioout{\Pi_1}{c} \cup \ioin{\Pi_2}{c}{d}}$,
  then $\alpha \in \tracesof{\iooutreveal{\Pi_1}{c} \cup
    \ioinadversary{\Pi_2}{d}}$. By
  Theorem~\ref{thm:basic.protocol-centric}, the claim holds in the
  latter trace set. Since it is protocol-centric, it must also hold in
  the former trace set. The theorem is therefore proven.
\end{proof}
\fi

When one protocol establishes session for some nonce run term, and a
second protocol expects a session key as input, we can use the nonce
run term from the first protocol as input to the second, without
compromising any protocol-centric security properties.

Note that in the following two results, we restrict to exactly one
protocol creating output and one protocol taking input. This is for
simplicity, and can easily be solved using protocol tags to create
many distinct variants of a single protocol.

\iffalse The reason is
that $\sessionuniqueclaim$ is too weak to be used when multiple
protocols create output, since it does not guarantee that two distinct
protocols create distinct outputs. The simple solution, creating a new
security notion suitable for this case would require several new
supporting theorems, since the security property would not be
protocol-centric. Since the following theorems are sufficient for many
applications, we leave this work to the future.

Proving a result for multiple protocols taking input (from one or more
protocols creating output) requires a different approach. Essentially,
we need to ensure that, when running the first protocol, every agent
intends to execute the same protocol after the initial protocol
completes. We solve this by creating multiple versions of the initial
protocol(s) using protocol tags, one for each protocol taking input.
This means that we do not require a new theorem for this case. \fi

\begin{thm}
  \label{thm:protocol-centric.session.passing}
  Let $\Pi_1$ and $\Pi_2$ be two protocol sets, such that $P \in
  \Pi_1$ is the only protocol establishing $c$ and $Q \in \Pi_2$ is
  the only protocol taking $c$ as input for $d$. Let $\thelabel$ be
  the label of a protocol-centric claim event in $\Pi_1$ or $\Pi_2$.
  If $\Pi_1$ and $\Pi_2$ are independent under the trace restrictions
  $\chi_{IO!}(\anything;P,Q,c,d)$ and
  $\chi_\secretclaim(\anything;Q,d)$, then
  \begin{equation*}
    \satisfies{\Pi_1}{\{\thelabel_i\} \cup \{\thelabel\}} \wedge
    \satisfies{\ioinsession{\Pi_2}{d}}{\{\thelabel_i'\} \cup \{\thelabel\}} \Rightarrow
  \satisfies{\ioout{\Pi_1}{c} \cup
    \ioinunique{\Pi_2}{c}{d}}{\thelabel} \text,
  \end{equation*}
  where $\{\thelabel_i\}$ are labels of claim events
  $\{\claimevent_{\thelabel_i}(r_i,\sessionclaim,c)\}$ in every role
  of $P$, and $\{\thelabel_i'\}$ are labels of claim events
  $\{\claimevent_{\thelabel_i'}(r_i',\secretclaim,d)\}$ in every role
  $Q$, occurring before any $\sendevent$ or $\readevent$ event in the
  role specification.
\end{thm}
\begin{proof}
  Let $\thelabel$ be in a protocol $R$. First, we note that the map
  $\tau: \tracesof{\ioout{\Pi_1}{c} \cup \ioinunique{\Pi_2}{c}{d}}
  \rightarrow \tracesof{\Pi_1 \cup \ioinsecret{\Pi_2}{d}}$ from
  (\ref{eq:replacing.input.by.secret}) on page~\pageref{eq:replacing.input.by.secret} exists, since the
  $\sessionclaim$ claims imply corresponding $\secretclaim$ claims.
  Since for some substitution $s$, $s(\pi_R(\alpha)) =
  \pi_R(\tau(\alpha))$, we only need to show that the $\sessionclaim$
  trace restriction is satisfied for the input to $Q$, and the result
  will follow from Theorem~\ref{thm:basic.protocol-centric}. Because
  of the $\dataagreeclaim$ claims, we have a full session for $P$, and
  by the $\sessionuniqueclaim$ claims, this session is unique. Thus
  for any $\stopevent$ event for any role $r$ of $P$, there are no
  other $\stopevent$ events for that role with the same value for $c$.
  Hence, the $\sessionclaim$ trace restriction is satisfied.
\end{proof}

Finally, we combine the previous theorems into a statement about
preservation of session secrecy.

\begin{cor}
  \label{cor:session.preserved.under.session.passing}
  Let $\Pi_1$ and $\Pi_2$ be two protocol sets, such that $P \in
  \Pi_1$ is the only protocol establishing $c$ and $Q \in \Pi_2$ is
  the only protocol taking $c$ as input for $d$. Let $\thelabel$ be
  the label of a (weak) session claim in $\Pi_1$ or $\Pi_2$. If
  $\Pi_1$ and $\Pi_2$ are independent under the trace restrictions
  $\chi_{IO!}(\anything;P,Q,c,d)$ and
  $\chi_\secretclaim(\anything;Q,d)$, then
  \begin{equation*}
    \satisfies{\Pi_1}{\{\thelabel_i\} \cup \{\thelabel\}} \wedge
    \satisfies{\ioinsession{\Pi_2}{d}}{\{\thelabel_i'\} \cup
      \{\thelabel\}} \Rightarrow \satisfies{\ioout{\Pi_1}{c} \cup
    \ioinunique{\Pi_2}{c}{d}}{\thelabel} \text,
  \end{equation*}
  where $\{\thelabel_i\}$ are labels of claim events
  $\{\claimevent_{\thelabel_i}(r_i,\sessionclaim,c)\}$ in every role
  of $P$, and $\{\thelabel_i'\}$ are labels of claim events
  $\{\claimevent_{\thelabel_i'}(r_i',\secretclaim,d)\}$ in every role
  $Q$, occurring before any $\sendevent$ or $\readevent$ event in the
  role specification.
\end{cor}
\begin{proof}
  First we apply Theorem~\ref{thm:basic.secret} and
  Theorem~\ref{thm:basic.protocol-centric} to establish
  $\satisfies{\Pi_1 \cup \ioinsession{\Pi_2}{d}}{\{\thelabel_i\} \cup
    \{\thelabel_i'\}}$ (separately establishing the three parts of the
  session claims: secret, session-unique and data-agree). The results
  follows by further applications of
  Theorem~\ref{thm:secret.secret.passing} and
  Theorem~\ref{thm:protocol-centric.session.passing}.
\end{proof}

\subsection{Composition}

In this section we study sequential composition and show how
certain security properties of a composed protocol follow from
security properties of the subprotocols analyzed in a
multi-protocol setting.

As discussed in Section \ref{sec:introduction}, sequential composition
(without passing information) of two protocols does not in general
preserve synchronisation. The problem is that if there is no mechanism
to bind the two subprotocols to each other in the composed
protocol, different runs can be interleaved with each other, breaking
synchronisation. Therefore, there must be a mechanism that connects
the two subprotocols. We achieve this by letting the first subprotocol
pass information to the next subprotocol. (A slightly more general
definition of chaining composition allowing more than one parameter
appears in PCL.)

\begin{defn}
  Let $rs_1 = (\elist_1\cdot \stopevent(r),\idtyper_1)$ and
  $rs_2 = (\createevent(r) \cdot\ $ %
  $ elist_2,\idtyper_2) \in
  \rolespec$. The \emph{sequential composition of role specifications} $rs_1$ and
  $rs_2$ is the role specification $rs_1\cdot rs_2=(\elist_1\cdot\elist_2,
  \idtyper)$ where
  \begin{equation*}
  \idtyper(x)  = \left\{
      \begin{array}{ll}
        \idtyper_1(x) &  \text{\ if\ } \idtyper_1(x) \text{\ is defined;} \\
        \idtyper_2(x) &  \text{\ if\ } \idtyper_1(x) \text{\ is undefined and\ } \idtyper_2(x) \text{\ is
        defined;}\\
        \text{undefined} & \text{\ otherwise.}
      \end{array}
  \right.
\end{equation*}
\end{defn}

\begin{defn}
  Let $P$ and $Q$ be two protocols such that $\domain(P) =
  \domain(Q)$, $\idset(P) \cap \idset(Q) = \emptyset$. If $P$
  establishes $c$, and $d$ is a parameter in all roles of $Q$, the
  \emph{chaining composition} $\chaincomp{P}{Q}{c}{d}$ of $P$ and $Q$
  is defined as:
  \begin{equation*}
    \chaincomp{P}{Q}{c}{d} \defeq
    \bigl\{
    r \mapsto P(r) \cdot Q(r)[c/d]
    \bigm|
    r \in \domain(P)
    \bigr\} \text,
  \end{equation*}
  where $Q(r)[c/d]$ denotes replacing $d$ by $c$ in the role
  specification $Q(r)$.
\end{defn}

Note that every event is relabeled after this composition, but there
is a natural correspondence between the labels of $P$ and $Q$, and the
labels of $\chaincomp{P}{Q}{c}{d}$ (excluding the $\stopevent$ event
of $P$ and $\createevent$ event of $Q$).

The formal model described in Section~\ref{sec:framework} allows us to
define explicitly the concept of passing information from one protocol
to another. This exactly coincides with the idea of input and output,
modeled using parameters. However, simply passing information does
not suffice to preserve synchronization. Intuitively, if agents of the
first subprotocol do not all share the value to be passed on to the
next subprotocol, a mismatch between different runs of the agents may
occur and synchronization may be broken. Likewise, the next
subprotocol must ensure that all agents got passed the same value. In
order to ensure this, we use data agreement for the value passed
between subprotocols.

\begin{thm}
  \label{thm:composition.synch}
  Let $P$, $Q$ be protocols such that $P$ establishes $c$, $d$ is a
  parameter in all roles of $Q$ and $\chaincomp{P}{Q}{c}{d}$ is
  defined. Let $\Pi$ be a set of protocols, $P,Q \not\in \Pi$. Let
  $\{\thelabel_i\}$ be a set of labels for one injective
  synchronization claim event and one data agreement claim event with
  argument $c$ in every role of $P$, the synchronization claim events
  appearing after all $\readevent$ and $\sendevent$ events in the
  role. Let $\thelabel$ and $\thelabel'$ be the labels of a data
  agreement claim event with $d$ as argument and an (injective)
  synchronization claim event, respectively, in some role $Q$ such
  that the claim event labelled $\thelabel$ causally precedes the one
  labelled $\thelabel'$. Let $\thelabel''$ be the label of a
  corresponding injective synchronization claim event in
  $\chaincomp{P}{Q}{c}{d}$. Then
  \begin{equation*}
    \satisfies{\{\ioout{P}{c},
      \ioinunique{Q}{c}{d}\} \cup
      \Pi}{\{\thelabel_i\}\cup\{\thelabel,\thelabel'\}}
    \Rightarrow \satisfies{\{\chaincomp{P}{Q}{c}{d}\} \cup \Pi }{
      \thelabel'' } \text.
  \end{equation*}
\end{thm}

\begin{proof}
  Let $\alpha\in \tracesof{\{\chaincomp{P}{Q}{c}{d}\} \cup \Pi}$. We
  map $\alpha$ to a trace $\alpha' \in \tracesof{\{\ioout{P}{c},
    \ioinunique{Q}{c}{d}\} \cup \Pi}$ as follows: Without loss of
  generality we can assume that the set $\runidset$ of run identifiers
  is a subset of non-negative integers. Let the highest occurring run
  identifier in $\alpha$ be $rid_B$, and suppose we have a run with
  run identifier $rid$ of a role $r$ of $\chaincomp{P}{Q}{c}{d}$ with
  events $\alpha_{i_1},\alpha_{i_2},\dots,\alpha_{i_k}$. If there is
  no event in the run corresponding to an event of the protocol $Q$,
  we relabel all of the events to be events of $P$. Otherwise, let
  $i_l$ be the first event in the run corresponding to an event of
  $Q$.
  \begin{enumerate}
  \item We relabel $\alpha_{i_1},\dots,\alpha_{i_{l-1}}$ to be events
    of $P$, and $\alpha_{i_l},\dots,\alpha_{i_k}$ to be events of $Q$.
  \item We change the $\sigma$-parts of the instantiations of
    $\alpha_{i_l},\dots,\alpha_{i_k}$ so that they are only defined at
    identifiers of $Q$.
  \item We change the run identifier of
    $\alpha_{i_l},\dots,\alpha_{i_k}$ to $rid + rid_B + 1$.
  \item \begin{sloppypar} We apply to every other event in the trace the substitution
    $\{ x\markid{rid} \mapsto x\markid{rid + rid_B + 1} \mid x \in
    \idset(Q) \wedge \idtyper(x) = \consttype \}$, where $Q(r) =
    (\elist,\idtyper)$.\end{sloppypar}
  \item We insert a $\stopevent$ for $P(r)$ and a create event $Q(r)$
    with the proper value for the parameter $d$ just before
    $\alpha_{i_l}$ in the trace.
  \end{enumerate}
  When this operation is performed for every run of a role of
  $\chaincomp{P}{Q}{c}{d}$, we get a trace $\alpha' \in
  \tracesof{\{\ioout{P}{c}, \ioinunique{Q}{c}{d}\} \cup \Pi}$, and
  this gives us a map
  \begin{equation}
    \label{eq:splitting.compositions}
    \tau: \tracesof{\{\chaincomp{P}{Q}{c}{d}\} \cup \Pi} \rightarrow
    \tracesof{\{\ioout{P}{c}, \ioinunique{Q}{c}{d}\} \cup \Pi} \text.
  \end{equation}

  Now consider a run of role $r$ with run identifier $rid_r$ where the
  claim event with label $\thelabel''$ occurs. First, we note that
  because of injective synchronization for $P$, we have a unique cast
  $\cast$ for $P$ in $\alpha'$. This translates into a potential cast
  $\cast'$ for $Q$ in $\alpha'$ given by $\cast'(r') = \cast(r) +
  rid_B + 1$, as well as a cast for $\chaincomp{P}{Q}{c}{d}$ in
  $\alpha$. By data agreement for $c$ in $P$, we know that every run
  in the cast agree on the value of $c$. Since $P$ does not take any
  input, the value of $c$ must originate with one of the roles, hence
  it must also be unique among all the runs of $P$. Further, by data
  agreement on $d$ in $Q$, we know that $\cast'$ really is a cast for
  $Q$ in $\alpha'$, it is unique, that every member of the cast agrees on the value
  of $d$, and that this value is the same as the value of $c$.

  Now we verify the $\isynchclaim$ claim with label $\thelabel''$ for
  $rid_r$ in $\alpha$ with the unique cast $\cast$. Consider a role
  $r'$ and a label $\thelabel_x$ such that $\thelabel_x \in
  \chaincomp{P}{Q}{c}{c}(r)$ and $\thelabel_x \in
  \chaincomp{P}{Q}{c}{c}(r')$. We must show that there are two events
  with this label in $\alpha$, belonging to the cast, and sending and
  reading the same message. Because of synchronization in
  $\alpha'$, we find matching events belonging to the cast for the
  corresponding labels in $\alpha'$. Note that since the map $\tau$
  only changes instantiations by applying a substitution, if the
  content of the messages is the same in $\alpha'$, the same must hold
  in $\alpha$. We conclude that the injective synchronization claim
  with label $\thelabel''$ is satisfied in $\alpha$.
\end{proof}

The following theorem states the conditions under which secrecy is preserved in a sequential protocol composition.

\begin{thm}
  \label{thm:composition.secret}
  Let $P$, $Q$ be protocols such that $P$ establishes $c$, $d$ is a
  parameter in all roles of $Q$ and $\chaincomp{P}{Q}{c}{d}$ is
  defined. Let $\Pi$ be a set of protocols, $P,Q \not\in \Pi$. Let
  $\thelabel$ be the label of a $\secretclaim$ claim event in $P$ or
  $Q$, and let $\thelabel'$ be the corresponding label in
  $\chaincomp{P}{Q}{c}{d}$. Then
  \begin{equation*}
    \satisfies{\{\ioout{P}{c}, \ioinunique{Q}{c}{d}\} \cup
      \Pi}{\thelabel}
    \Rightarrow \satisfies{\{\chaincomp{P}{Q}{c}{d}\} \cup \Pi }{
      \thelabel' } \text.
  \end{equation*}
\end{thm}
\begin{proof}
  We use the map $\tau$ from \eqref{eq:splitting.compositions}. Note
  that instantiations are changed by at most a nonce run term renaming
  under this map, so the intruder's knowledge is also changed by at
  most a nonce run term renaming. The predicate derived from the
  secret claim does not change its value under nonce run term
  renaming, from which the result follows.
\end{proof}

The same conditions that preserve secrecy, also preserve $\sessionuniqueclaim$ and $\dataagreeclaim$ in a sequential protocol composition.

\begin{thm}
  \label{thm:composition.session-unique.data-agreement}
  Let $P$, $Q$ be protocols such that $P$ establishes $c$, $d$ is a
  parameter in all roles of $Q$ and $\chaincomp{P}{Q}{c}{d}$ is
  defined. Let $\Pi$ be a set of protocols, $P,Q \not\in \Pi$. Let
  $\thelabel$ be the label of a $\sessionuniqueclaim$ or
  $\dataagreeclaim$ claim event in $P$ or $Q$, and let $\thelabel'$ be
  the corresponding label in $\chaincomp{P}{Q}{c}{d}$. Then
  \begin{equation*}
    \satisfies{\{\ioout{P}{c}, \ioinunique{Q}{c}{d}\} \cup
      \Pi}{\thelabel}
    \Rightarrow \satisfies{\{\chaincomp{P}{Q}{c}{d}\} \cup \Pi }{
      \thelabel' } \text.
  \end{equation*}
\end{thm}
\begin{proof}
  We use the map $\tau$ from \eqref{eq:splitting.compositions}. By the
  construction of the map, if some value only appears in one claim
  event in $\tau(\alpha)$, it will only appear in one claim event in
  $\alpha$ as well. This proves the theorem for $\sessionuniqueclaim$.

  As for data agreement, if events for some cast exist in
  $\tau(\alpha)$ with the correct value, they will also exist in
  $\alpha$. This proves the theorem for $\dataagreeclaim$.
\end{proof}

By definition of $\sessionclaim$ and $\wsessionclaim$,
the preceding two theorems give conditions under which these two properties are preserved.
This is expressed in the following

\begin{cor}
  \label{cor:composition.session}
  Let $P$, $Q$ be protocols such that $P$ establishes $c$, $d$ is a
  parameter in all roles of $Q$ and $\chaincomp{P}{Q}{c}{d}$ is
  defined. Let $\Pi$ be a set of protocols, $P,Q \not\in \Pi$. Let
  $\thelabel$ be the label of a (weak) $\sessionclaim$ claim event in
  $P$ or $Q$, and let $\thelabel'$ be the corresponding label in
  $\chaincomp{P}{Q}{c}{d}$. Then
  \begin{equation*}
    \satisfies{\{\ioout{P}{c}, \ioinunique{Q}{c}{d}\} \cup
      \Pi}{\thelabel}
    \Rightarrow \satisfies{\{\chaincomp{P}{Q}{c}{d}\} \cup \Pi }{
      \thelabel' } \text.
  \end{equation*}
\end{cor}
\begin{proof}
  We use Theorem~\ref{thm:composition.secret} and
  Theorem~\ref{thm:composition.session-unique.data-agreement} to
  establish that the requisite $\secretclaim$, $\sessionuniqueclaim$ and
  $\dataagreeclaim$ claims hold, and the result follows.
\end{proof}

\iffalse
\note{SM}{The use of labels l, l', l'' in the theorems requires from
the user that he has to map these to particular (types of) claims.
Would it be an idea to make them mnemonic through indexing, e.g.
$l_{isynch}$ ?}
\note{SM}{I already thought while typing this suggestion that it would
be an excellent improvement in the revision of our paper and an
excellent way to introduce new errors right before the deadline :-)}
\note{CC}{Self-evaluation saves the day once again.}
\fi

\section{Mobile \wimax}
\label{sec:wimax}

In this section we apply our framework to a handful of protocols
from the security sublayer of the IEEE 802.16-2005
amendment~\cite{wimaxieee} of the IEEE 802.16-2004
standard~\cite{wimaxieee-2004}, commonly and in the following
referred to as (mobile) \wimax. The aim is not a complete
verification of \wimax as this would constitute a research topic
of its own. Instead, we use \wimax to illustrate our framework on
a real-world protocol suite and as a measure for our progress
towards the goal of a comprehensive theory of protocol
verification. We stress that even in the limited setting we
consider, our methods are strong enough to draw useful conclusions
about protocol design and security flaws.

\subsection{Introduction}\label{sec:wimax.introduction}

The IEEE 802.16-2004 standard specifies the air interface of fixed
broadband wireless access systems supporting multimedia services
in local and metropolitan area networks. The 802.16-2005 amendment
addresses mobility of subscriber stations and features new security protocols.

A brief overview of the communication between a mobile station and base station follows.
The communication
starts at the mobile station's network entry with the \ranging{}
protocol. Its purpose is to set up physical communication
parameters and assign a basic connection identifier to the
requesting mobile station. This protocol is periodically executed
later to re-communicate the physical communication parameters.
Next, a \registration{} protocol is carried out in order to allow
the mobile station into the network. During this protocol, the
base station and mobile station's security capabilities are
negotiated. The base station and mobile station can agree on
unilateral or mutual authentication, or no authentication at all,
and on a variety of key management protocols. The key management
protocols are periodically repeated to update the traffic
encryption keys. The
entire authentication chain is repeated on a less frequent basis. Once
the traffic encryption keys are established, user data protocols
start. To avoid service interruptions, traffic encryption keys have
overlapping lifetimes.

The authentication and key management protocols are specified in
the security sublayer of \wimax. The security sublayer is meant to
provide subscribers with privacy and authentication and operators
with strong protection from theft of
service~\cite[Chapter 7]{wimaxieee}. %
It employs an authenticated client/server key management protocol
in which the base station controls distribution of keying material
to the mobile station.

The security sublayer consists of two component protocols, an
encapsulation protocol for securing packet data across the network
and a key management protocol
 providing the secure distribution of
keying data from the base station to the mobile station. In the
following sections we will focus on the key management protocol.
Through this key management protocol, the base station and mobile
station are to synchronize keying data and the base station is
meant to use the protocol to enforce conditional access to network
services.

The overall security goals mentioned in the specification are ``no
theft of service'' for the operator of the base station and
``confidentiality'' for the user of the mobile
station. Confidentiality in \wimax is defined to be ``privacy'' and
``authenticity''\cite[Chapter~7, footnote~6]{wimaxieee}. The
specification gives vague ideas for the security properties that
the subprotocols have, by calling them for instance
authentication protocols or key update protocols. But for a
thorough security analysis, precise security claims need to be
made for each subprotocol and the relation between the security
properties achieved by the subprotocols and the overall security
goals needs to be understood. We are addressing these issues in
the following section.

\subsection{Key management in the Security Sublayer}\label{sec:wimax.intro.key.management}

The privacy key management (PKM) component of the security
sublayer consists of authentication and key establishment
protocols. Depending on the negotiated security capabilities, the
IEEE-802.16-2004 PKM protocols or the new PKM version 2 protocols will
be executed. In the following security analysis, we consider the
sequence of the three PKM version 2 protocols \pkmvrsa,
  \pkmvsatek, and \pkmvtek{}. \pkmvrsa{} authenticates
the base station (\bs) and mobile station (\ms) and establishes a
shared secret which is used by \pkmvsatek{} and \pkmvtek{} to
secure the exchange of traffic encryption keys (TEKs). \wimax does
not explicitly state what the security claims of these three
protocols are. As indicated in the Introduction, it is stated that
the sequential composition of the three protocols achieves strong
authentication and privacy for the mobile station, and strongly
protects the base station from theft of service. Furthermore, it
is implicitly stated that the established keys are shared secrets
and \pkmvrsa{} is called a mutual authentication protocol.

We are making these properties more precise by imposing the
following requirements on the composition of the three protocols.
In order to provide ``strong protection against theft of
service''~\cite[Chapter 7]{wimaxieee} for the base station,
the client station has to
be strongly authenticated at the end of the protocol composition, i.e. the role of the base station has to satisfy the $\isynchclaim$ claim and all key
material must be secret. Note that if the $\isynchclaim$ claim is true at the
end of the protocol composition then we are guaranteed that every
message read by $\bs$ up to that point
has been sent by $\ms$ and exactly matches the message sent by
$\ms$. Thus no theft of service can have occurred up to that point.
In order to provide privacy and authenticity for the subscriber
station, we demand that the base station is strongly
authenticated at the end of the protocol composition,
and that all session keys and symmetric encryption keys are secret and
unique.

We argue now that the following security properties for the three
protocols and their sequential composition need to be fulfilled and prove in
the next section that these properties indeed imply our set
security goals. Since \pkmvrsa{} needs to authenticate \ms{} and
\bs{} and establish a shared secret which is to be used as a key
later on, it has to satisfy $\isynchclaim$ for both roles and
$\sessionclaim$ for the shared secret. Since \pkmvsatek{} and
\pkmvtek{} need to establish further keys while keeping up the
authentication property between \ms{} and \bs{}, they both need to
satisfy $\nisynchclaim$ for both roles, $\sessionclaim$ for the
shared secret, and at least $\wsessionclaim$ for the traffic
encryption keys\footnote{Since $\wsessionclaim$ and $\dataagreeclaim$
  imply $\sessionclaim$, the use of the $\tek$'s in user data
  protocols by both roles
  will automatically imply the $\sessionclaim$ claim at that point.}. This implies in
particular that the shared secret and traffic encryption keys have
to be secret in all three protocols. We summarize these requirements
in Table~\ref{tbl:summary.secprop}. It corresponds to the summary of verified
security properties in Section~\ref{s:composition}.

\begin{table}[!ht]
\begin{center}
\begin{tabular}{|l|l|}
\hline
Protocol & Security Properties\\
\hline
\pkmvrsa & $\isynchclaim, \sessionclaim(\prepak)$\\
\pkmvsatek & $\isynchclaim,\sessionclaim(\prepak),\wsessionclaim(\tek)$\\
\pkmvtek & $\nisynchclaim,\sessionclaim(\prepak),\wsessionclaim(\tek)$\\
\hline
\end{tabular}
\end{center}
\caption{Required security properties for the PKMv2 subprotocols in \wimax.}
\label{tbl:summary.secprop}
\end{table}

Before we start the description of the protocols,
some technical remarks on our \wimax model are in order.
We will restrict ourselves to non-handover scenarios
with unicast communication, leaving out multicast and broadcast
communication, mesh communication, and group security.
We further restrict ourselves to one Security Association as
opposed to
  a list of several security associations offered by the base station. This
 is simply for convenience as it implies that there is only one pair of TEK keys instead of several
 pairs identified by SAID's and managed by parallel sessions of the
 \pkmvtek{} protocol.

In our model, we have simplified messages by omitting irrelevant terms and
headers. In particular, the fact that all messages in \wimax are
formatted using a type/length/value (TLV) scheme, we show only implicitly
by giving appropriate names to
identifiers. Entries in a TLV list are called attributes.
\wimax specification states that both roles silently
discard all messages that do not contain all required
attributes and skip over unknown attributes.
Note that the use of TLVs together with signatures and hashed message authentication codes in \wimax also implies that we may disregard type flaw attacks.
Finally, we model hash functions as encryptions with a special public key, known by all parties, whose inverse key is not known by anyone. We thus write
$\enc{m}{h(\mathit{salt})}$ for the message $m$ to which a
message authentication code has been attached. Private key signatures will be indicated by $\enc{m}{\skms}$ and $\enc{m}{\skbs}$.

\subsubsection{\pkmvrsa}

The $\pkmvrsa$ protocol (Figure~\ref{fig:pkmv2rsa}) is the initial mutual authentication
protocol. It is repeated periodically to update the $\prepak$. Its
purpose is to establish a shared secret $\prepak$ (called pre-PAK
in \wimax) between $\ms$ and $\bs$. The shared secret is used to
derive the authentication key from which the keys for hashed
message authentication codes and symmetric encryptions are derived.

According to specification, the \rsareq{} message consists of
\msrand, \mscert, \said{} and a signature over a SHA-1 hash of
these fields using \ms's secret key, whose corresponding public key
\bs{} learns from \mscert.
We model this by sending \mscert{} outside of
$\enc{\ldots}{\skms}$. The same applies to \bscert{} in the
\rsareply{} and \rsareject{} messages.

\mscwimaxsettings
\setlength{\envinstdist}{3.1cm}
\setlength{\instdist}{8.5cm}
\begin{figure}[!htb]
\centering{ {\small
\begin{msc}{PKMv2-RSA}

\declinst{ms}{$\macert, \mscert, \pkca$}{$\ms$}
\setlength{\envinstdist}{1.7cm}  %
\declinst{bs}{$\bscert, \pkca$}{$\bs$}

\gate[l][c]{$\authinfo$}{ms}
\mess{$\macert$}{ms}{bs}
\nextlevel
\action{$\msrand$}{ms}
\nextlevel[1.2]
\gate[l][c]{$\rsareq$}{ms}
\mess{$\mscert,\enc{\msrand, \said, \mscert}{\skms}$}{ms}{bs}
\nextlevel
\setlength{\actionwidth}{3cm}
\action{$\bsrand,\prepak$}{bs}
\nextlevel
\inlinestart[1.5cm][1cm]{choice}{or}{ms}{bs}
\nextlevel
\gate[l][c]{$\rsareply$}{ms}
\mess{$\substack{\ \\\bscert,\enc{\msrand, \bsrand,
\enc{\prepak,\ms}{\pkms},\bscert}{\skbs}}$}{bs}{ms}
\nextlevel
\inlineseparator{choice}
\nextlevel
\gate[l][c]{$\rsareject$}{ms}
\mess{$\bscert, \enc{\msrand, \bsrand,\bscert}{\skbs}$}{bs}{ms}
\nextlevel
\inlineend{choice}
\nextlevel
\gate[l][c]{$\rsaack$}{ms}
\mess{$\enc{\bsrand}{\skms}$}{ms}{bs}
\end{msc}
}}

\bigskip

\begin{tabular}{|p{0.2\linewidth}|p{0.7\linewidth}|}
\hline
\macert & manufacturer's certificate\\
\mscert & mobile station's certificate $\enc{\ms, \pkms}{\skman}$\\
\bscert & base station's certificate $\enc{\bs, \pkbs}{\skca}$\\
\said & security association Id, equal to $\bcid$\\
\msrand & mobile station's nonce\\
\bsrand & base station's nonce\\
\prepak & preliminary primary authentication key, \\
        & actually another nonce created by the base station\\

\hline
\end{tabular}
\caption{The $\pkmvrsa$ protocol.} \label{fig:pkmv2rsa}
\end{figure}

\subsubsection{\pkmvsatek}

The $\pkmvsatek$ protocol (Figure~\ref{fig:pkmv2satek}) is a
three-way handshake protocol which follows the $\pkmvrsa$
protocol. Its purpose is to update the traffic encryption keys if
they already exist. The protocol's messages are authenticated by
hashing them with keys derived from $\prepak$
established by the $\pkmvrsa$ protocol. Different keys are used
for uplink and downlink traffic. The base station sends a
challenge to the mobile station, which the mobile station repeats
in its request for updated key material and thus proves liveness
and knowledge of the shared secret (by using the derived $\hmac$
key) to the base station. The base station answers then with
updated key material, repeating the mobile station's nonce.

\setlength{\envinstdist}{2.3cm}
\begin{figure}[!htb]
\centering{ {\small
\begin{msc}{PKMv2 SA-TEK}

\declinst{ms}{$\akid,\tekzero,\tekone$}{$\ms$}
\declinst{bs}{$\akid',\tekzero,\tekone$}{$\bs$}

\action{$\bsrand'$}{bs}
\nextlevel
\gate[l][c]{$\satekchall$}{ms}
\mess{$\enc{\bsrand', \akid' }{h(\hmacd)}$}{bs}{ms}
\nextlevel
\action{$\msrand'$}{ms}
\nextlevel[1.2]
\gate[l][c]{$\satekrequest$}{ms}
\mess{$\enc{\msrand', \bsrand', \akid }{h(\hmacu)}$}{ms}{bs}
\nextlevel
\gate[l][c]{$\satekresponse$}{ms}
\mess{$\substack{\ \\\enc{\msrand', \bsrand',
  \akid,\enc{\tekzero',\tekone'}{\kek}}{h(\hmacd)}}$}{bs}{ms}
\end{msc}
}}

\bigskip

\begin{tabular}{|p{0.2\linewidth}|p{0.7\linewidth}|}
\hline \pak & Primary authentication key derived from $\prepak$ as
follows:
\\
    & $\pak = keyedhash(\prepak, \ms \concat \bs)$. \\
\ak & Authentication key derived from $\pak$ as follows: \\
    & $\ak = keyedhash(\pak, \ms \concat \bs \concat \pak)$. \\
\kek & key encryption key derived from $\ak$ as follows: \\
     & $\hmacu \concat \hmacd \concat \kek =
        keyedhash(\ak, \ms \concat \bs )$. \\
\hmacd & HMAC key derived from $\ak$ for authenticating downlink communication\\
\hmacu & HMAC key derived from $\ak$ for authenticating uplink
communication\\
$\msrand'$ & mobile station's nonce\\
$\bsrand'$ & base station's nonce\\
\aksn & \ak\ sequence number, essentially a 2 bit counter\\
\akid & \ak\ Id: $\akid=keyedhash(AK,AKSN\concat \ms\concat \bs)$\\
$\akid'$ & \ak\ Id of new AK if re-authenticating\\
\hline
\end{tabular}

\caption{The $\pkmvsatek$ protocol.} \label{fig:pkmv2satek}
\end{figure}

\subsubsection{\pkmvtek}

The $\pkmvtek$ protocol (Figure~\ref{fig:pkmv2tek}) allows the
mobile station to obtain the most recent $\tek$ key from the base
station.

\setlength{\envinstdist}{2.3cm}
\begin{figure}[!htb]
\centering{ {\small
\begin{msc}{PKMv2 Key}

\declinst{ms}{\tekzero,\tekone}{$\ms$}
\setlength{\envinstdist}{1.7cm}  %
\declinst{bs}{\tekzero,\tekone}{$\bs$}

\action{$\nonce{}$}{ms}
\nextlevel
\gate[l][c]{$\tekrequest$}{ms}
\mess{$\enc{\nonce{} }{h(\hmacu)}$}{ms}{bs}
\nextlevel

\inlinestart[1.5cm][1cm]{choice}{or}{ms}{bs}
\nextlevel

\gate[l][c]{$\tekreply$}{ms}
\mess{$\enc{\enc{\tekzero',\tekone'}{\kek}, \nonce{} }{h(\hmacd)}$}{bs}{ms}

\nextlevel
\inlineseparator{choice}
\nextlevel

\gate[l][c]{$\tekreject$}{ms}
\mess{$\enc{\nonce{} }{h(\hmacd)}$}{bs}{ms}

\nextlevel
\inlineend{choice}
\nextlevel

\end{msc}
}}

\bigskip

\begin{tabular}{|p{0.2\linewidth}|p{0.7\linewidth}|}
\hline
\tekzero & Older traffic encryption key \\
\tekone & Newer traffic encryption key \\
$\tekzero'$, $\tekone'$ & Updated traffic encryption keys, replacing \tekzero, \tekone, respectively\footnote{In a normal run \tekzero' is equal to \tekone, while \tekone' has been freshly generated by \bs.}\\
\kek & Key Encryption Key (see above) \\
\hline
\end{tabular}

\caption{The $\pkmvtek$ protocol.} \label{fig:pkmv2tek}
\end{figure}

\subsection{Applying the Framework}

We begin by analyzing the three protocols described in the
previous section in isolation and then we apply our theorems to
their sequential composition. To facilitate later exposition, we will, during
the course of the analysis, simplify the protocols presented
above. Since the aim of this work is not a careful and formal
analysis of these short subprotocols, we will reason on an
informal level for clarity, backed up by the automated
verification tool Scyther. After that, we will analyse the
composition in detail.

\subsubsection{Analysis of \pkmvrsa}
\label{sec:rsaanalysis}

We analyze the $\pkmvrsa$ protocol without the \authinfo{} message
which according to specification is only being sent right after
\ranging{} and never again. We first consider $\isynchclaim$ for
\ms{} and \bs{} and then the $\sessionclaim$ property for \prepak.

\paragraph{$\isynchclaim$.} Since we have a choice in the third message between \rsareply{}
and \rsareject{}, we will first analyze the branch with \rsareply,
then the branch with \rsareject.

\begin{description}
\item[\rsareply.] Note that the structure of \rsareq, \rsareply, \rsaack{} is
similar to the standard X.509 protocol, except for the last
message where the identity of \bs{} is missing. As a consequence,
it suffers from a man-in-the-middle attack causing \bs{} to not
synchronize. The intruder executes the man-in-the-middle attack by
taking advantage of an \ms{} trying to connect to the intruder and
redirecting those messages to a \bs. In~\cite{XuHu06} the authors
describe a similar, but slightly more complicated attack, in which the
intruder uses two runs of \ms{} to impersonate \ms{} to \bs.
Both attacks can be found using Scyther.

The attacker can not, however, impersonate \bs{} to \ms, and
the \ms{} role in fact
still synchronizes and the \prepak{} remains secret. It is also
interesting to note that \rsareq, \rsareply, \rsaack{} followed by
\pkmvsatek{} has agreement (an authentication property slightly weaker
than synchronisation, see\cite{CrMaVi06}) for both \ms{} and \bs{}. However, we
still consider the lack of \bs{}'s identity in \rsaack{} a design
flaw, since it is supposed to be a ``mutual authentication''
protocol according to specification and hence, in its current
form, breaks a modular design principle: small changes in other
protocols (for instance \pkmvsatek) could break the security of
the entire composition.

For future reference, we write \pkmvrsax{} to denote the protocol
consisting of \rsareq, \rsareply, and \rsaack' where \rsaack' is
the message $\enc{\bsrand, \bs}{\skms}$ from \ms{} to \bs.
Furthermore, we will denote the \prepak{} in \pkmvrsax{} simply by
\prepakx. \pkmvrsax{} has injective synchronization for both
roles.

\item[\rsareject.] As in the branch analyzed above, the structure
\rsareq, \rsareject, \rsaack{} resembles the X.509 standard. Here
however, both \rsareject{} and \rsaack{} are missing the recipient's
ID. Therefore neither \bs{} nor \ms{} synchronize (the weaker
agreement notion is not satisfied either).
A \prepak{} is not being sent, thus the
secrecy claim is void. The fact that \ms{} does not synchronize
can be abused for a denial of service attack.
\end{description}

We amend the flaws pointed out above by
considering $\pkmvrsax$ instead of %
$\z{PKMv2}$- $\z{RSA}$. Note the absence of
a $\rsareject$ message from $\pkmvrsax$. A \rsareject{} message
sent from \bs{} to \ms{} terminates the protocol. In order for
\ms{} to communicate with \bs{} it has to start over with the
\ranging{} protocol. For this reason, we will simplify the analysis
of the composition in Section~\ref{s:composition}, without
affecting any of the security properties we are interested in, if
we consider the protocol without the \rsareject{} message; instead
of a send event corresponding to the \rsareject{} message, the run
of the \bs{} role ends.

The specification of protocol $\pkmvrsax$ is given below. For brevity,
we omit the typing of identifiers and use the shorthand for the
message contents as displayed in Figure~\ref{fig:pkmv2rsa}.
The descriptive labels of the claim events are inserted for further
reference.

$$
  \begin{array}{lll}
    \pkmvrsax(\ms) & = &
      \createevent_{\pkmvrsax 1} (\ms) \seq
      \sendevent_{\pkmvrsax 2} (\ms, \bs, \rsareq) \seq
      \readevent_{\pkmvrsax 3} (\bs, \ms, \rsareply ) \seq \\
      && \sendevent_{\pkmvrsax 4} (\ms, \bs, \enc{\bsrand, \bs}{\skms}) \seq
      \claimevent_{\labelpmsisynch}(\ms,\isynchclaim) \seq \\
      && \claimevent_{\labelpmssession}(\ms,\sessionclaim,\prepakx) \seq
      \stopevent_{\pkmvrsax 5} (\ms)
      \\
    \pkmvrsax(\bs) & = &
      \createevent_{\pkmvrsax 6} (\bs) \seq
      \readevent_{\pkmvrsax 2} (\ms, \bs, \rsareq) \seq
      \sendevent_{\pkmvrsax 3} (\bs, \ms, \rsareply ) \seq \\
      && \readevent_{\pkmvrsax 4} (\ms, \bs, \enc{\bsrand, \bs}{\skms}) \seq
      \claimevent_{\labelpbsisynch}(\bs,\isynchclaim) \seq \\
      && \claimevent_{\labelpbssession}(\bs,\sessionclaim,\prepakx) \seq
      \stopevent_{\pkmvrsax 7} (\bs)
      \\
\end{array}
$$

In the remainder, we will use the abbreviation $\labelpisynch$ to
stand for the combination of $\labelpmsisynch$ and $\labelpbsisynch$.
We will use similar abbreviations for the other claims and protocols.

Using Scyther, we prove synchronisation. Given the fact that
$\pkmvrsax$ satisfies the \emph{loop-property} from~\cite{CrMaVi06},
we establish
\begin{equation*}
  \satisfies{\{\pkmvrsax\}}{\{\labelpisynch \}} \text.
\end{equation*}

\paragraph{$\sessionclaim$.}

We use Scyther to verify that $\prepak$ is secret. The 
$\z{data}$- $\z{agree}$
property is satisfied because of injective
synchronization, and the fact that $\prepak$ is part of a message
causally preceding the synchronization claims. Finally,
$\sessionuniqueclaim$ is also satisfied because of injective synchronization and the fact that
$\prepak$ is a constant in one of the roles appearing only in one send event,
accompanied, within a signature, by the recipient's nonce.

The established result is
\begin{equation*}
  \satisfies{\{\pkmvrsax\}}{\{\labelpsession \}} \text.
\end{equation*}

\subsubsection{Analysis of \pkmvsatek}

We call \pkmvsatekx{} the protocol obtained from \pkmvsatek{} by
introducing the parameter $\hmacx$, which will obtain its value from the
constant $\prepakx$ produced by protocol $\pkmvrsax$. The
parameter $\hmacx$ will hold the shared secret established in
$\pkmvrsax$ from which the $\hmac$ and $\kek$ keys are derived.
Furthermore, the
collection of \tek{} keys will be denoted by $\tekx$ in
\pkmvsatekx. %
Thus, $\pkmvsatekx$ is up to renaming of constants equivalent
to $\pkmvsatek$.

We insert session claim events for $\hmacx$, weak session claim
events for $\tekx$, and injective synchronization at the end of
both roles, with labels
$\labelqmssessiond$,
$\labelqbssessiond$,
$\labelqmswsessione$,
$\labelqbswsessione$,
$\labelqmsisynch$, and
$\labelqbsisynch$.
This yields the following description of protocol $\pkmvsatekx$
(assuming $h1$, $h2$, and $h3$ are distinct hash functions).

$$
  \begin{array}{lll}
    \pkmvsatekx(\ms) & = &
      \createevent_{\pkmvsatekx 1} (\ms) \seq
      \readevent_{\pkmvsatekx 2} (\bs, \ms, \enc{\bsrand', \akid'
    }{h1(\hmacx)}) \seq \\
      && \sendevent_{\pkmvsatekx 3} (\ms, \bs, \enc{\msrand',
      \bsrand', \akid }{h2(\hmacx)}) \seq \\
      && \readevent_{\pkmvsatekx 4} (\bs, \ms, \enc{\msrand', \bsrand',
      \akid,\enc{\tekx}{h3(\hmacx)}} {h1(\hmacx)}) \seq \\
      && \claimevent_{\labelqmsisynch}(\ms,\isynchclaim) \seq
      \claimevent_{\labelqmssessiond}(\ms,\sessionclaim,\hmacx) \seq \\
      && \claimevent_{\labelqmswsessione}(\ms,\wsessionclaim,\tekx) \seq
      \stopevent_{\pkmvsatekx 5} (\ms)
      \\
    \pkmvsatekx(\bs) & = &
      \createevent_{\pkmvsatekx 6} (\bs) \seq
      \sendevent_{\pkmvsatekx 2} (\bs, \ms, \enc{\bsrand', \akid'
    }{h1(\hmacx)}) \seq \\
      && \readevent_{\pkmvsatekx 3} (\ms, \bs, \enc{\msrand',
      \bsrand', \akid }{h2(\hmacx)}) \seq \\
      && \sendevent_{\pkmvsatekx 4} (\bs, \ms, \enc{\msrand', \bsrand',
      \akid,\enc{\tekx}{h3(\hmacx)}} {h1(\hmacx)}) \seq \\
      && \claimevent_{\labelqbsisynch}(\bs,\isynchclaim) \seq
      \claimevent_{\labelqbssessiond}(\bs,\sessionclaim,\hmacx) \seq \\
      && \claimevent_{\labelqbswsessione}(\bs,\wsessionclaim,\tekx) \seq
      \stopevent_{\pkmvsatekx 7} (\bs)
      \\
\end{array}
$$

Again, we verify injective
synchronization and secrecy for $\hmacx$ and $\tekx$ using
Scyther. Note that the verification has to be done for the
trace restriction $\ioinsession{\pkmvsatekx}{d}$.
Session trace restrictions can be simulated in Scyther using technical
tricks, but are expected to be supported natively in the future.

The $\sessionuniqueclaim$ property follows from arguments analogous to the ones shown for $\prepakx$ in $\pkmvrsax$ and the session trace restriction for both $\hmacx$ and $\tekx$.
Data agreement for $\hmacx$ follows from injective synchronization and the appearance of $\hmacx$ in a message causally preceding the $\isynchclaim$ claim for both roles.
For $\tekx$ we do not
get data agreement for the $\bs$ role, since $\tekx$ is sent in the
last message for which $\bs$ has no guarantee that $\ms$ received it.

We have shown
\begin{equation*}
  \satisfies{\{\ioinsession{\pkmvsatekx}{\hmacx}\}}{\{
  \labelqsessiond,
  \labelqwsessione,
  \labelqisynch
  \}}
\end{equation*}

\subsubsection{Analysis of \pkmvtek}

Similarly to the previous two protocols, we will let $\pkmvtekx$ denote the
protocol obtained from $\pkmvtek$ by denoting the \hmac{} keys by
$\hmacx$, the old $\tek$ keys by $\tekx$ and the new $\tek$ keys
by $\tekx'$. We let $\pkmvtekx$ only consist of the $\tekrequest$ and
$\tekreply$ messages, since the alternative has exactly the same security
properties.

We have $\isynchclaim$ for the $\ms$ role, shown by applying Scyther and using the loop property, but only $\nisynchclaim$ for $\bs$. %

The $\sessionclaim$ property for $\hmacx$ and $\wsessionclaim$
property for $\tekx'$ can be shown in exactly the same manner as
for the $\pkmvsatekx$ protocol.

$$
  \begin{array}{lll}
    \pkmvtekx(\ms) & = &
      \createevent_{\pkmvtekx 1} (\ms) \seq
      \sendevent_{\pkmvtekx 2} (\ms, \bs, \enc{\nonce{} }{h2(\hmacx)}) \seq \\
      && \readevent_{\pkmvtekx 3} (\bs, \ms,
      \enc{\enc{\tekx'}{h3(\hmacx)}, \nonce{}}
      {h1(\hmacx)}) \seq \\
      && \claimevent_{\labelrmsisynch}(\ms,\isynchclaim) \seq
      \claimevent_{\labelrmssessiond}(\ms,\sessionclaim,\hmacx) \seq \\
      && \claimevent_{\labelrmswsessione}(\ms,\wsessionclaim,{\tekx'}) \seq
      \stopevent_{\pkmvtekx 5} (\ms)
      \\
    \pkmvtekx(\bs) & = &
      \createevent_{\pkmvtekx 6} (\bs) \seq
      \readevent_{\pkmvtekx 2} (\ms, \bs, \enc{\nonce{} }{h2(\hmacx)}) \seq \\
      && \sendevent_{\pkmvtekx 3} (\bs, \ms,
      \enc{\enc{\tekx'}{h3(\hmacx)}, \nonce{}}
      {h1(\hmacx)}) \seq \\
      && \claimevent_{\labelrbssynch}(\bs,\nisynchclaim) \seq
      \claimevent_{\labelrbssessiond}(\bs,\sessionclaim,\hmacx) \seq \\
      && \claimevent_{\labelrbswsessione}(\bs,\wsessionclaim,{\tekx'}) \seq
      \stopevent_{\pkmvtekx 7} (\bs)
      \\
\end{array}
$$

Abbreviating the claim labels, we obtain
\begin{align*}
  \satisfies{\{\ioinsession{\pkmvtekx}{\hmacx}\}}{\{&
\labelrmsisynch,
\labelrbssynch,\\
&\labelrsessiond,
\labelrwsessione
\}} \text.
\end{align*}

\subsubsection{The composition}
\label{s:composition}

In the preceding three subsections we have established that
\begin{align}
 & \satisfies{\{\pkmvrsax\}}{\{\labelpisynch,\labelpsession \}}\label{eq:Pprops} \\
 &  \satisfies{\{\ioinsession{\pkmvsatekx}{\hmacx}\}}{\{
  \labelqsessiond,
  \labelqwsessione,
  \labelqisynch
  \}}\label{eq:Qprops}\\
 &
 \begin{aligned}
\satisfies{\{\ioinsession{\pkmvtekx}{\hmacx}\}}{\{&
  \labelrmsisynch,
  \labelrbssynch, \\ &
  \labelrsessiond,
  \labelrwsessione
  \}}
 \end{aligned}
\label{eq:Rprops}
\end{align}

The methodology for the verification
of these facts is standard. In what follows, we apply the
collection of theorems in our framework to show how the
established properties imply correctness of the composed protocol
$P\cdot Q\cdot R$.

Note that \pkmvrsax, \pkmvsatekx, and \pkmvtekx{}  are mutually
strongly independent since the messages, as stated in
Section~\ref{sec:wimax.intro.key.management}, are TLV encoded,
the signatures or hashed message authentication codes
are made over the entire message,
 and the message structures are
different in the three protocols. More precisely, protocol $P$ is
independent from $Q$
and $R$ because all messages in $P$ are signed by private keys and
hence the signatures will not be accepted by either role in protocols
$Q$ and $R$, their messages being authenticated using the shared secret
keys. Protocols $Q$ and $R$ are strongly independent, since they don't
have a message in common in which all required attributes are
identical.

Using strong independence, we can now deduce that \pkmvrsax{} followed
by \pkmvsatekx{} satisfies injective synchronization, session, and
weak session as follows.

 By
Theorem~\ref{thm:protocol-centric.session.passing} and equations \eqref{eq:Pprops} and \eqref{eq:Qprops},  we can preserve the
$\isynchclaim$ property for $\pkmvrsax$, $\pkmvsatekx$. %
By Corollary~\ref{cor:session.preserved.under.session.passing}, $\prepakx$
and $\hmacx$ keep the $\sessionclaim$ property, and $\tekx$ the $\wsessionclaim$ property.

Therefore, we obtain
\begin{align*}
  \satisfies{\{\ioout{\pkmvrsax}{\prepakx},\ioin{\pkmvsatekx}{\prepakx}{\hmacx}\}}{\{& \labelpisynch,\labelqisynch,\\
    & \labelpsession,\labelqsessiond,\labelqwsessione\}}
  \text,
\end{align*}

and using Theorems~\ref{thm:composition.synch} (to obtain
$\isynchclaim$), %
and Corollary~\ref{cor:composition.session}
($\wsessionclaim$ and $\sessionclaim$)
we get
\begin{equation}\label{eq:PQ}
\satisfies{\{\pkmvrsax\cdot\pkmvsatekx\}}{\{
\labelpqisynch, \labelpqsessionc, \labelpqwsessione \}}.
\end{equation}
Here we assume that in the composed protocol $\pkmvrsax\cdot\pkmvsatekx$
roles $\bs$ and $\ms$ are extended with appropriate claim events.
We use the three labels $\labelpqisynch$, $\labelpqsessionc$,
and $\labelpqwsessione$ to refer to these claims.

Next, by Theorem~\ref{thm:protocol-centric.session.passing}
we establish from
\eqref{eq:Rprops} and \eqref{eq:PQ} injective synchronization
for both roles and by
Corollary~\ref{cor:session.preserved.under.session.passing}
$\sessionclaim$ for $\prepakx$ and $\wsessionclaim$ for $\tekx$:
\begin{align*}
\satisfies{
\{\ioout{\pkmvrsax\cdot\pkmvsatekx}{\prepakx},\ioin{\pkmvtekx}{\prepakx}{\hmacx}\}
}
{\{& \labelpqisynch, \\
  & \labelpqsessionc,%
  \labelpqwsessione, \\
  & \labelrmsisynch,
  \labelrbssynch,\\
  & \labelrsessiond,
  \labelrwsessione
\}}
\end{align*}

Using Theorem~\ref{thm:composition.synch} and
Corollary~\ref{cor:composition.session} one more time,
we can show that the entire composition satisfies
injective synchronization and (weak) session secrecy for the shared secret $\prepakx$ and the traffic encryption key $\tekx$:
\begin{align*}
\satisfies{\pkmvrsax\cdot\pkmvsatekx\cdot\pkmvtekx}{\{ & \labelpqrisynch,
  \labelpqrsessionc, \\ & \labelpqrwsessione, \labelpqrwsessionep
  \}}
\end{align*}

These are exactly the overall security properties we have formulated in
Section~\ref{sec:wimax.intro.key.management}. As we have shown in
that section, these security properties are a precise
interpretation of the security goals stated in the \wimax
specification.

\subsection{\wimax: Conclusion and related work}

We have verified that the sequential composition of the
key management protocols \pkmvrsa,
\pkmvsatek, and \pkmvtek{} satisfies strong authentication and session
secrecy for keys derived from the shared secret and for traffic encryption
keys. However, in order to achieve this verification, we had to first
formulate precise security properties for the subprotocols and their
composition, based on our interpretation
of the rather vague security goals specified in \wimax.

While we have shown that our strong authentication property
$\isynchclaim$ holds for the protocol composition as stated, it has to be noted
that a composition consisting of repeated iterations of the $\pkmvtek$
protocol would fail to satisfy this property, due to replay
attacks. In practice, such attacks would only be a nuisance, not a
security threat, since $\secretclaim$ would still hold true for all keys, and
such an active intruder would not be able to learn anything more than
a passive, listening, one. Thus, in future work, the \wimax protocols
will be analyzed with an appropriately weakened notion of authentication.

Although several studies on \wimax security have appeared,
none of these offer a precise and compositional verification of the
\wimax key management protocols. Closest to our work is a preliminary
study in~\cite{Ka06}, which sketches the steps towards a compositional
verification.
An analysis of some protocols from the 2001 version of the \wimax
standard is conducted in~\cite{JohnstonW04}. Their main observation is
that the old protocols achieve unilateral authentication, while mutual
authentication of $\bs$ and $\ms$ is required. The proposed fixes
clearly found their way into the current standard. However, the
protocols are studied in isolation and the authors did not apply
formal reasoning to prove the fixed protocols correct.
Xu et al.~\cite{XuMaHu06,XuHu06} analyzed several isolated protocols
from the current \wimax standard. Through informal reasoning, they
discovered the attack mentioned in Section~\ref{sec:rsaanalysis} and
proposed a fix. The modified protocol is proved correct by using the
BAN-logic~\cite{BAN96}, which is known to be incomplete with respect
to insider attacks.
Finally, we mention an analysis of the current \wimax standard using
the TLA+ logic in~\cite{XuMaHu06}.
The authors study the
composition of the three mentioned key management protocols as one
single protocol. Exploiting symmetry reduction techniques, they manage to
apply the TLC model checker to verify the composed protocol. However,
their focus is not on the key management properties that we
investigate (such as authentication and secrecy), but on detecting
a class of denial-of-service attacks. In order to validate liveness,
they focus on the state machines underlying the protocols.

\section{Related Work}
\label{sec:relatedwork}

In this section we address work related to the composition of security
protocols.
\iffalse
Within the area of security protocol analysis, the problem of
compositionality has been a known problem for several years,
e.g.~\cite{heintze96comp}. Initial results stated that security
properties are in general not compositional, as
in~\cite{kelsey97protocol,Alves-Foss99a}.  Later, compositionality has been described
as one of the open challenges for security protocol analysis
in~\cite{meadows01open,Cr04a}.
Early work on identifying and addressing the problem include
\cite{DBLP:journals/jcs/Paulson98,DBLP:conf/csfw/DurginMP01,mitchell03comp}.
Nevertheless, the vast majority of
formalisms and tools have only addressed single-protocol
(i.e.~non-composed) verification. An initial attempt within the Strand
Spaces approach~\cite{thayer99mixed} has led to some theoretical
results about compositionality.
\note{SR}{Text up to here has been taken into the introduction}
\fi
We discuss strand spaces, as well as the more recent Protocol Composition
Logic, in some detail below. Afterwards we address
related theoretical results, and discuss some attempts at the
verification of composed protocols.

\subsection{Strand Spaces}

Within a modified version of the Strand Spaces
framework~\cite{thayer98}, called the Mixed Strand Spaces
model~\cite{thayer99mixed}, some results about compositionality
have been proven.  In~\cite{guttman00protocol}, a disjoint
encryption theorem is proven. This theorem states that if two
protocols have sufficiently different encrypted messages (at the
trace/run level), composing them in parallel will not introduce new
attacks.
In terms of methodology, this work is closely related to ours:
given {\em any} two correct protocols, what abstract properties
should they satisfy, in order to ensure that their composition is
correct?

The main differences between their approach and ours, are that (1)
they only consider parallel composition, and (2) verification
of the disjoint encryption property has to be done at the level of
traces. With respect to the first item, we note that the
approach does not allow for the decomposition of large sequential
protocols into smaller ones, as can be done with the chaining
theorems presented here.
Similarly, there are no concepts such as
session-secrecy. Consequently, the disjoint encryption approach
cannot be used for the compositional verification of strongly
dependent subprotocols such as those present in \wimax, for instance.

The second item represents a more significant drawback of the approach.
The verification of disjoint encryption has to be
performed at the trace level of the composed protocols. For some
protocols, this can be easily, but nevertheless manually, deduced from
the protocol specification, but in many other cases (e.g.~where session
keys are used in protocols) the only way to verify that the disjoint
encryption property holds is by inspecting the traces of the composed
protocol. Therefore, in such cases there is no expected improvement on
the more traditional approach of, for example, model checking all traces of the
composed protocol.

As a more subtle drawback, the proof given for the disjoint
encryption theorem assumes that the security properties do not
include ordering constraints. Thus, it is not immediately
possible
to apply the theorem for the verification of strong authentication
properties, such as e.g.~the synchronization property.

One advantage of the approach is that their results also hold
for protocols that include tickets, something that we have
explicitly excluded here.

\subsection{Protocol Composition Logic}

\iffalse
One of the significant recent developments in protocol analysis is the
development of Protocol Composition Logic (PCL) \cite{DDMR07}. This is
a logic designed to reason about protocol correctness.
%
%
%
%
PCL has been applied in a number of
case studies, including verifying the TLS and IEEE 802.11i protocols
\cite{DBLP:conf/ccs/HeSDDM05} and contract signing protocols
\cite{DBLP:journals/tcs/BackesDDMT06}.

\note{SR}{The above paragraph has been taken into the introduction}
\fi

%
%
One of the most significant theoretical results in the work on
Protocol Composition Logic (PCL) \cite{DDMR07} is a strategy for dealing with
protocol composition. The basic idea is to prove the protocols correct
in isolation by constructing a correctness proof in the logic. Certain
so-called invariants are then identified in the correctness proofs,
such that protocol correctness follows from these invariants only. If
these invariants are not violated by the other protocols, it is an
easy consequence that correctness is retained under composition, since
correctness follows from the invariants alone.

In contrast to this very general strategy, our composition theorems
identify specific classes of protocols that can be composed in certain
ways. The advantage of our approach is that it is highly amenable to
automatic verification (as demonstrated in Section~\ref{sec:wimax}),
especially when combined with the trivially verifiable strong
independence property. 
In contrast, the full generality of proof derivation in PCL seems difficult to automate.

It is easy to see that our notion of strong independence is a
stronger requirement than the invariants used for
composition in PCL. Nevertheless, strong independence is trivial
to verify, hence highly suited for an automatic verification strategy.
It is possible that composition theorems similar to ours, based on
strong independence, could be recovered in the PCL framework.

The PCL invariant approach can
deal with cases that our notion of independence cannot. The
reason for this is that
independence only considers ciphertext terms and their
origins, while PCL invariants cover more general statements.
Conversely, since independence is verified for traces while PCL
invariants are verified over so-called basic sequences, it is not
immediately obvious that the PCL approach can deal with every case our
independence notion can deal with. Investigating this relationship is
an interesting topic for future research.

We believe that many ideas and techniques used in PCL can be reused in
our framework. The techniques used to identify and verify invariants
could possibly be used to prove independence. Due to the highly manual
nature of the PCL compositionality strategy, we expect any such work
to be complementary to the theory developed in this paper, to be used
only when automatic techniques fail.

An alternative approach is taken in the PDa tool~\cite{PavlovicD:ARSPA06}. In
this tool, %
an axiomatic theory is set up
to reason about protocol refinement and composition. The tool uses ideas from
PCL to reason about invariants. The tool can provide automatic discharging of
simple proofs. However, the user has to provide sufficiently strengthened
invariants to allow for compositional proofs. The axiomatic theory does not
have a notion of run (or process or thread), similar to e.g.~BAN logic, and as a result
only very weak notions of authentication can be considered.

\subsection{Related theoretical results}

The complex problem of compositionality has been approached from a variety of
angles. Many of these approaches are restricted to weak forms of authentication, such as
\cite{Bugliesi04,mpauthentication,BugliesiFM04}. When only such weaker forms are considered, compositionality
results can be achieved on the basis of simpler challenge-response
mechanisms within a protocol, similar to the authentication tests
from~\cite{guttman01authentication}. The existence of these mechanisms in a
protocol does not ensure synchronization, or even agreement.

Other approaches have considered secrecy, e.g.~\cite{Ju01}. Here a
notion of secrecy is defined within the context of stream-processing
functions. Using the notion of an $m$-secrecy protecting process, a
result is given that states that two such processes can be safely
composed. Furthermore, it is shown that such a process remains
secrecy-protecting under refinement. Similar to the Strand Spaces
approach, it is unclear how one can establish that a process (or
protocol) satisfies the required conditions for the stated theorems.

In the area of information flow analysis, which is related to the secrecy-only
approach, there are a number of results and supporting tools,
e.g.~\cite{mantel02comp,heintze96comp,Ju01,Focardi96b,Focardi97}. However,
because of fundamental differences in the underlying models, these results
cannot be used for the compositional analysis of security protocols such as
\wimax.

In~\cite{canetti-environmental} the observation is made that the correctness of
security protocols depends on the assumptions on the environment. In the wrong
environment, or in the context of specific protocols, seemingly secure
protocols are incorrect. The authors give no specific conditions or properties.

Over the past decades compositional verification has received quite
some attention from the process algebra community and was applied
successfully in the verification of complex concurrent systems (for
an overview of these techniques, see e.g.~\cite{Roever01}). However,
these techniques do not seem to carry over easily to the process
algebras developed especially for security protocols, such as the
spi calculus~\cite{Abadi99}. An attempt has been made
in~\cite{Boreale02}. Here, compositionality is interpreted as a
congruence property of a bisimulation-like relation over several
process operators. Although the authors provide compositional rules
for (restricted) parallel and action-prefix composition, rules for
general sequential composition are absent, making it impossible to
apply this work to e.g.\ the \wimax protocol suite. Moreover, the rule
for parallel composition poses a very strong restriction on the set
of  processes that may run in parallel with any given process.

Furthermore, the security properties treated are secrecy and weak
forms of authentication. It is not obvious how general
protocol-centric properties and especially injectivity can be
expressed by means of the bisimulation relation provided. Finally,
we note that the proposed methodology has severe limitations with
respect to the verification of actual protocols. As an example, the
authors prove correctness of a version of the Wide Mouthed Frog
protocol, which is obviously insecure in the standard setting for
security protocols. This problem is due to the fact that their
theory only supports the verification of fixed scenarios.

\subsection{Verification of composed security protocols}

In the area of protocol verification, it seems that the first attempt at
verification of parallel subprotocols was made
in~\cite{meadowsdidanalysis}, where the interaction between subprotocols
was investigated manually.

An attempt at composing security protocol proofs within a theorem-proving
environment was made in~\cite{sheyner00}. In this work, the authors
construct compositional proofs for a specific protocol, in order to
work towards a general theory. The conclusion of the authors is that even for a
single protocol, the approach requires much manual work, and that scaling
problems might cause this approach to be infeasible.

More recently fully automated verification of composed protocols was performed
in~\cite{Cr06} employing the same basic framework and tool used here.
However, this verification, too, has been limited to protocols that are
composed in parallel.

\section{Conclusion}
\label{sec:conclusion}

We make two significant contributions in this paper. First, we create
a framework for easy verification of a large and useful class of
security protocols built from smaller subprotocols. Second, we
initiate a study of \wimax, by applying our framework to the
composition of three protocols from its security sublayer.
This is done by first verifying that the three protocols are independent and then analyzing each subprotocol in isolation. The results of this analysis are used to deduce properties
about the composition of the subprotocols, consisting of the
subprotocols running in parallel with suitable transfer of information.
Consequently, this allows us to derive properties of the composed
protocol.

We do not claim that our framework can deal with every possible
security protocol. One important restriction is the requirement in
many theorems that subprotocols are independent. This makes it
difficult to use our framework for analysis of protocols that do not
naturally split into independent subprotocols. As we have argued, protocol tags are
a reasonably cost-effective way to design protocols that are amenable
to analysis in our framework. \wimax is just one example of
protocols in which such techniques are in use today. We believe this is a
very reasonable approach to future protocol design.

A significant advantage of our framework is ease of use. If we
consider the \wimax analysis, the Scyther tool automatically
proves secrecy and synchronization for the subprotocols. Since
session-uniqueness and data agreement claims have not yet been
implemented in Scyther, a small amount of reasoning is needed to
prove that the protocols have these properties in isolation. Once
the properties are established, however, using the theorems to
deduce the security properties of the composed protocol is
essentially trivial. As the \wimax analysis to some degree shows,
it should be possible to verify protocols without an intimate
knowledge of the underlying semantics described in
Section~\ref{sec:framework}, since a tool like Scyther (once it is
suitably extended) can deal with the proofs needed at this level.

An interesting feature of our framework is that the theorem statements
are not strongly connected to the underlying semantics. They are
therefore in a sense independent of the semantics. Indeed, we believe
the framework could be transferred to any other semantics powerful enough to
express at least the notion of independence and the security
properties, and which has a similar execution model.

In general, our theorems are tight in the sense that if any
precondition is relaxed, the theorem is no longer true. Of course,
some theorems could be extended in natural ways, and other theorems
have many specialized variations. For the current work, we believe
such extensions would add little value. Instead, such results should
be proved as needed, slowly increasing the knowledge about how
composition works.

In this work, we have defined the protocol-centric class of security
properties and proved many theorems for that class. Likewise, we can
define other classes of properties, for instance properties that only
consider the intruder's memory. Studying such classes of properties
and proving theorems about them is an interesting future topic.

Another useful contribution in this paper is our definition of
protocol independence. Currently, we have only described one way to
achieve independence, namely protocol tags. There are several other
ways one could imagine achieving independence, for instance through
some notion of separate key infrastructures. One can also imagine
other notions of independence that allow general theorems to be
proved. Such notions would create new protocol design strategies and
allow more protocols to be analyzed. We intend to continue our
work on this topic.

The requirement in our semantics that variables only contain nonce run
terms prevents us from expressing protocols using tickets
in the semantics. The requirement is
only essential for Theorem~\ref{thm:strong.independence}. A
more significant problem is the fact that security properties such as
synchronization or agreement do not make sense in the context of
tickets, since some roles are by definition insensitive to the content
of the tickets. An important topic for future work will be to extend
our framework with new security properties and new theorems for
ticket-based security protocols.

In our framework we discuss how to compose protocols. While sequential
composition is the natural notion of protocol composition, there are
other possible composition operators that are natural to discuss, such
as the choice operator allowing one out of two protocols to run.
Extending our framework with such operators and theorems to support
reasoning with them is an important future topic.

As we have already noted, the Scyther tool does not have support for
every security property we have defined, nor for every trace
restriction. In the near future, we intend to extend Scyther with
support for these security properties and trace restrictions. A
related task is the creation of a new tool to formally verify reasoning in
our framework. Essentially, this tool will use
Scyther as a back-end to analyze the subprotocols, then it will verify that every theorem
application is valid. This will allow automated verification of large
protocols. As the body of theorems in our framework increases, so will the
power of the tool when the theorems are added.

We have analyzed  the security requirements of \wimax{} and shown that
a somewhat restricted variant of the protocol satisfies these
requirements, all by reasoning in our framework and analyzing small
subprotocols in isolation.
We believe our study, though not complete, is a useful
first step towards a complete analysis of the security requirements of
\wimax, as well as towards a verification of the entire protocol suite. In the
future, we intend to work out a complete analysis of the security
sublayer of \wimax.

Today, most new protocols are not verified (in any sense of the word)
when they are released, for example, as standards. We believe this is because
today, verification of any sizable protocol is the exclusive province
of the few skilled specialists and researchers working in the area. An
important goal of current research is to remedy this problem. As the
analysis of the \wimax protocols show, our work is a significant
first step towards a framework for security protocol analysis (with
tool support) that could be used by engineers to verify protocols
during design, allowing a proper security analysis of the protocol before
release.

\paragraph*{Acknowledgment}
We thank Eric Kaasenbrood for his help in understanding and
modeling \wimax.
We also thank the anonymous reviewers whose comments have
helped to improve this paper.

\bibliographystyle{plain}
\bibliography{report}

\end{document}